\documentclass[11pt]{article}
\usepackage[utf8]{inputenc} 	
\usepackage[T1]{fontenc}
\usepackage{lmodern}
\usepackage{yhmath}

\usepackage{amsthm,amsmath,amssymb,amsbsy,bbm,mathrsfs,supertabular,
eurosym,graphicx,enumitem,xcolor}
\usepackage{mathtools}

\newtheoremstyle{BBstyle0}  {}{}{\itshape}{}{\bfseries}{}{6pt}{}
\newtheoremstyle{BBstyle1}  {3pt}{3pt}{\rmfamily}{}{\itshape}{: }{3pt}{}
\newtheoremstyle{BBstyle2}  {3pt}{3pt}{\itshape}{}{\bfseries\large}{}{0pt}{}
\newtheoremstyle{BBstyle3}  {}{}{\itshape}{}{\bfseries}{: }{3pt}{}
\newtheoremstyle{BBstyle4}  {}{}{\rmfamily}{}{\bfseries}{}{6pt}{}
  
\usepackage[normalem]{ulem} 

\usepackage[authoryear]{natbib}
\newtheorem{lem}{Lemma}
\newtheorem{prop}{Proposition}

\theoremstyle{definition}

\usepackage[english]{babel}

\newcommand{\pa}[1]{\left({#1}\right)}

\newcommand{\cro}[1]{\left[{#1}\right]}
\newcommand{\ab}[1]{\left|{#1}\right|}
\newcommand{\ac}[1]{\left\{{#1}\right\}}

\newcommand{\argmin}{\mathop{\rm argmin}}

\newcommand{\E}{{\mathbb{E}}}

\newcommand{\N}{{\mathbb{N}}}
\renewcommand{\P}{{\mathbb{P}}}
 
\newcommand{\R}{{\mathbb{R}}}

\newcommand{\sF}{{\mathscr{F}}}
\newcommand{\sG}{{\mathscr{G}}}

\newcommand{\sQ}{{\mathscr{Q}}} 
\newcommand{\sR}{{\mathscr{R}}}

\newcommand{\sW}{{\mathscr{W}}}
\newcommand{\sX}{{\mathscr{X}}}
\newcommand{\sY}{{\mathscr{Y}}}

\DeclareMathAlphabet{\mathscrbf}{OMS}{mdugm}{b}{n}

\newcommand{\sbB}{{\mathscrbf{B}}}

\newcommand{\sbP}{{\mathscrbf{P}}}
\newcommand{\sbQ}{{\mathscrbf{Q}}} 
\newcommand{\sbR}{{\mathscrbf{R}}}

\newcommand{\cA}{{\mathcal{A}}}
\newcommand{\cB}{{\mathcal{B}}}

\newcommand{\cE}{{\mathcal{E}}}

\newcommand{\cI}{{\mathcal{I}}}

\newcommand{\cM}{{\mathcal{M}}}

\newcommand{\cO}{{\mathcal{O}}}

\newcommand{\cR}{{\mathcal{R}}}

\newcommand{\cW}{{\mathcal{W}}}
\newcommand{\cX}{{\mathcal{X}}}
\newcommand{\cY}{{\mathcal{Y}}} 
\newcommand{\cZ}{{\mathcal{Z}}}

\newcommand{\gh}{{\mathbf{h}}}

\newcommand{\gr}{{\mathbf{r}}}

\newcommand{\gw}{{\mathbf{w}}}

\newcommand{\gP}{{\mathbf{P}}}
 
\newcommand{\gR}{{\mathbf{R}}}
 
\newcommand{\gT}{{\mathbf{T}}}

\newcommand{\gX}{{\mathbf{X}}}
 
\newcommand{\gZ}{{\mathbf{Z}}}

\newcommand{\bs}[1]{\boldsymbol{#1}}

\newcommand{\bsG}{{\bs{G}}}

\newcommand{\bsX}{{\bs{X}}}
\newcommand{\bsY}{{\bs{Y}}} 

\newcommand{\ggamma}{\bs{\gamma}}
\newcommand{\gGamma}{\bs{\Gamma}}

\newlist{lista}{enumerate}{1}
\setlist[lista,1]{label=\alph*),ref=\alph*)}

\newlist{listi}{enumerate}{1}
\setlist[listi,1]{label=(\roman*),ref=(\roman*),align=left}

\newcommand{\eref}[1]{(\ref{#1})}
\usepackage[left=3cm,right=3cm, top=4.2cm, bottom=4.2cm]{geometry}
\usepackage[utf8]{inputenc}
\usepackage[T1]{fontenc}
\usepackage{yhmath}
\usepackage{amsthm,amsmath,amssymb,amsbsy,bbm,mathrsfs,supertabular,eurosym,enumitem,xcolor}
\usepackage{bm}
\usepackage{colortbl}
\usepackage{mathtools}
\usepackage{longtable}
\RequirePackage[colorlinks,citecolor=blue,urlcolor=blue]{hyperref}
\usepackage[]{natbib}
\usepackage{subcaption} 
\usepackage{graphicx} 
\usepackage{silence}
\makeatletter
\newcommand{\ssymbol}[1]{^{\@fnsymbol{#1}}}
\makeatother
\newtheoremstyle{BBstyle0}  {}{}{\itshape}{}{\bfseries}{}{6pt}{}
\newtheoremstyle{BBstyle1}  {3pt}{3pt}{\rmfamily}{}{\itshape}{: }{3pt}{}
\newtheoremstyle{BBstyle2}  {3pt}{3pt}{\itshape}{}{\bfseries\large}{}{0pt}{}
\newtheoremstyle{BBstyle3}  {}{}{\itshape}{}{\bfseries}{: }{3pt}{}
\newtheoremstyle{BBstyle4}  {}{}{\rmfamily}{}{\bfseries}{}{6pt}{}

\newtheorem{innercustomgeneric}{\customgenericname}
\providecommand{\customgenericname}{}
\newcommand{\newcustomtheorem}[2]{%
  \newenvironment{#1}[1]
  {%
   \renewcommand\customgenericname{#2}%
   \renewcommand\theinnercustomgeneric{##1}%
   \innercustomgeneric
  }
  {\endinnercustomgeneric}
}
\newcustomtheorem{customthm}{Proposition}

\title{Non-Data-Splitting Estimator Selection for Regression in Exponential Families}

\author{Juntong Chen\\[0.3cm]
Department of Applied Mathematics,\\[0.3cm] University of Twente}
\date{}

\newcommand{\gup}{\bs{\upsilon}}

\newcommand{\SMUCE}{\mathop{\rm SMUCE}}
\newcommand{\CCL}{\mathop{\rm CL}}

\def\gpen{\mathop{\rm \bf pen}\nolimits}

\def\bsg{{\ggamma}}
\def\bsG{{\overline{\bs{\Gamma}}}}

\def\bsg{{\mathbf{\gamma}}}

\renewcommand{\P}{{\mathbb{P}}}

\DeclareMathAlphabet{\mathscrbf}{OMS}{mdugm}{b}{n}

\def\gpen{\mathop{\rm \bf pen}\nolimits}

\def\bsg{{\ggamma}}
\def\bsG{{\overline{\bs{\Gamma}}}}

\newtheorem{theorem}{Theorem}

\newtheorem{corollary}{Corollary}

\newtheorem{assumption}{Assumption}
\theoremstyle{definition}
\newtheorem{definition}{Definition}

\renewcommand{\P}{{\mathbb{P}}}

\usepackage{algorithm}
\usepackage{algpseudocode}

\begin{document}
\maketitle

\begin{abstract}
We observe $n$ independent pairs of random variables $(W_{i}, Y_{i})$, where the conditional distribution of $Y_{i}$ given $W_{i}=w_{i}$ follows a one-parameter exponential family with parameter $\bsg^{*}(w_{i})\in\R$. Our goal is to estimate the regression function $\bsg^{*}$. We start with an arbitrary collection of piecewise constant candidate estimators based on our observations and, using the same data, select an estimator from this collection. Our approach is agnostic to the dependencies of the candidate estimators on the data, differing from methods like data splitting, cross-validation, and hold-out. To demonstrate its theoretical performance, we provide a non-asymptotic risk bound for the selected estimator. We then explain how to apply the procedure to changepoint detection in exponential families. The practical performance of the proposed approach is illustrated through a comparative simulation study under different scenarios and real datasets.
\end{abstract}

\section{Introduction}
We observe $n$ pairs of independent (but not necessarily i.i.d.) random variables, i.e. $X_{i}=(W_{i},Y_{i})$, for $i=1,\ldots,n$, with values in a measurable product space $(\sW\times\sY,\cW\otimes\cY)$. For each $i$, we assume that the conditional distribution of $Y_{i}$ given $W_{i}=w_{i}$ exists which we denote as $R_{i}^{*}(w_{i})$ and it belongs to a one-parameter exponential family with parameter $\bsg^{*}(w_{i})\in\R$. The aim of the present paper is to estimate the $n$ conditional distributions $R_{i}^{*}(w_{i})$ of $Y_{i}$ given $W_{i}=w_{i}$, i.e. to estimate the unknown function $\bsg^{*}$ on $\sW$, on the basis of the observations $\bsX=(X_{1},\ldots,X_{n})$.

Under this statistical setting, \cite{Baraud2020} introduced an estimation procedure based on single-model $\rho$-estimation (\cite{MR3595933}, \cite{BarBir2018}), where the risk bound for their estimator $\widehat{\bsg}$ is, up to a constant, the sum of an approximation term and a complexity term for the chosen model. This approach performs well when a suitable model for $\bsg^{*}$ is known in advance — one that approximates $\bsg^{*}$ effectively without being overly complex. However, designing such a model can be challenging with limited prior information. To address this, \cite{ChenModSel} proposed a model selection procedure, which mitigates the need for prior model knowledge but incurs high computational costs when dealing with a large number of models. This makes the model selection procedure primarily of theoretical interest. Developing an estimation strategy that balances accuracy and computational efficiency remains an important direction.

When $W_{i}=(i-1)/n$ (or $i/n$ in some literature) are deterministic and $\bsg^{*}$ is an unknown function on $[0,1)$ (or $(0,1]$), more work has been done in the literature. Here, $\bsY=(Y_{1},\ldots,Y_{n})\in\sY^{n}$ is observed, ordered by a covariate like time or genomic position. \cite{MR1859411} studied natural exponential families with quadratic variance functions, covering distributions like Gaussian, Poisson, and binomial, and proposed wavelet shrinkage estimators. \cite{MR1897045} extended this method to families with cubic variance functions. \cite{brown2010} also focused on quadratic variance families and recommended a mean-matching variance-stabilizing transformation to simplify the problem to homoscedastic Gaussian regression. In particular, when $\bsg^{*}:[0,1)\rightarrow I\subset \R$ is a right-continuous step function with an unknown number $N-1$ of changepoints ($N$ segments, $N\geq1$), the problem reduces to changepoint detection in exponential families. \cite{Frick2013MultiscaleCP} proposed the simultaneous multiscale changepoint estimator (SMUCE). They defined a multiscale statistic to evaluate the maximum local likelihood ratio over all discrete intervals where the estimator is constant. \citeauthor{refId0} (\citeyear{refId0}, \citeyear{cleynen2017}) used partitions from the pruned dynamic programming algorithm (\cite{rigaill2015pruned}) and proposed a penalized log-likelihood estimator, drawing on penalty construction by Birgé and Massart (\cite{MR1679028}, \cite{MR1462939}). They showed their estimator satisfies oracle inequalities. Both methods rely on maximum likelihood estimation and can infer extra changepoints to fit outliers while identifying true ones. Enhancing the stability of these estimators remains an open challenge.

Beyond methods tailored for changepoint detection in exponential families, detecting changes in random variable characteristics has a long history and has gained renewed interest in fields such as financial econometrics, climatology, bioinformatics, and signal processing. Recent theoretical advances include works by \cite{201011470} and \cite{10.1214/20-EJS1710} with a review provided in \cite{TRUONG2020107299}. Key methods include binary segmentation (BS) by \cite{10.2307/2529204}, circular binary segmentation (CBS) by \cite{PMID:15475419}, and pruned exact linear time (PELT) by \cite{Killick_2012}, which improves accuracy and efficiency. Wild binary segmentation (WBS) by \cite{Fryzlewicz2014} extends BS and is valued for its simplicity. Robust approaches like cumSeg (\cite{10.1093/bioinformatics/btq647}) and robseg (\cite{ChaDetectOut}) handle noise and outliers, while FDR (\cite{2016fdr}) controls false discovery rates for dense changepoints. Different methods often yield varying results, and no single method uniformly outperforms others (\cite{https://doi.org/10.1002/sta4.291}). Given a collection of (random) estimators, allowing the data to guide the selection process and automatically achieve near-optimal performance across cases is a compelling area for exploration.

This paper proposes a novel strategy to address the stated statistical problem: data-driven estimator selection (ES) without data splitting. Given observations $\bsX = (X_{1}, \ldots, X_{n})$, we consider a countable collection of piecewise constant (possibly random) candidate estimators for $\bsg^{*}$, denoted $\widehat\gGamma(\bsX)$, with potentially unknown dependence on $\bsX$. Our procedure compares these candidates pairwise using the same observations, enabling the data to identify the most suitable one. We establish a non-asymptotic risk bound for the selected estimator, comparing its risk to the infimum of risks over $\widehat\gGamma(\bsX)$. The ability to operate without knowing the dependencies of candidate estimators on the data sets our approach apart from methods like data splitting, cross-validation, and hold-out.

The paper is organized as follows. Section~\ref{statistical-setting} describes the statistical framework. Section~\ref{method} details our estimator selection procedure and its theoretical properties. In Section~\ref{apply-to-changepoint}, we apply the procedure to changepoint detection in exponential families. Section~\ref{simulation} presents a comparative simulation study showcasing the estimator's practical performance and stability. Section~\ref{realdata} exhibits the performance on two real datasets: DNA copy numbers and British coal disasters. All proofs from the main text, along with details of the test signals used in Section~\ref{simulation}, are provided in the Appendix.

\section{The statistical setting}\label{statistical-setting}
We consider $n$ pairs of independent (but not necessary i.i.d.) random variables $X_{i}=(W_{i},Y_{i})$, for $i\in\{1,\ldots,n\}$, taking values in a measurable product space $(\sX,\cX)=(\sW\times\sY,\cW\otimes\cY)$. Let $\sR$ denote the set of all probability measures on $(\sY,\cY)$, equipped with the Borel $\sigma$-algebra $\cR$ induced by the Hellinger distance (which shares the same topology as the total variation distance). Recall that the Hellinger distance between two probabilities $P=p\cdot\mu$ and $Q=q\cdot\mu$ dominated by a reference measure $\mu$ on a measurable space $(A,\cA)$ is defined as 
\begin{equation}\label{hellinger-distance-def}
h(P,Q)=\cro{\frac{1}{2}\int_{A}\pa{\sqrt{p}-\sqrt{q}}^{2}d\mu}^{1/2},
\end{equation}
a measure that is independent of the choice of the dominating reference measure $\mu$. For each $i$, we assume the conditional distribution of $Y_{i}$ given $W_{i}=w_{i}$ exists and is denoted $R_{i}^{*}(w_{i})$, where $R_{i}^{*}$ is a measurable function mapping $(\sW,\cW)$ to $\sR$. Notably, with the chosen $\cR$, for any $R\in\sR$ and any $i$, the mapping $w\mapsto h^{2}(R,R_{i}^{*}(w))$ on $(\sW,\cW)$ is measurable. We also recall the following definition.
 
\begin{definition}\label{general-exp}
Let $I$ be a non-trivial interval of $\R$ (i.e. $\mathring{I}\not=\emptyset$). We call a family of probabilities $\sQ_{0}=\left\{R_{\gamma},\;\gamma\in I\right\}$ on the measured space $(\sY,\cY,\mu)$ an exponential family under its general form, if $\sQ_{0}$ is a family of probabilities on $(\sY,\cY)$ admitting densities $\overline r_{\gamma}$ with respect to $\mu$ of the form, for all $\gamma\in I$
\begin{equation*}
\overline r_{\gamma}(y)=e^{u(\gamma)T(y)-A(\gamma)}h(y), \text{\quad for all\ }y\in\sY,
\end{equation*}
where $T$ is a real-valued measurable function on $(\sY,\cY)$ which does not coincide with a constant $\nu=h\cdot\mu$-a.e., $u$ is a continuous, strictly monotone function on $I$, $h$ is a nonnegative function on $\sY$ and $$A(\gamma)=\log\cro{\int_{\sY}e^{u(\gamma)T(y)}h(y)d\mu(y)}.$$
\end{definition}
Definition~\ref{general-exp} encompasses several important distributions, including Gaussian (with known variance), Poisson, binomial (with a fixed number of trials, such as Bernoulli), and gamma (with a fixed shape parameter, such as exponential). For simplicity, we rewrite $\sQ_{0}$ as $\left\{R_{\gamma}=r_{\gamma}\cdot\nu,\; \gamma\in I\right\}$, with the notation
\begin{equation}\label{general-den-2}
 r_{\gamma}(y)=e^{u(\gamma)T(y)-A(\gamma)}, \text{\quad for all\ }y\in\sY \text{\ and\ }\gamma\in I.
 \end{equation} 

The statistical framework considered here is broader than that presented in the introduction. Specifically, given the observations $\bsX=(X_{1},\ldots,X_{n})$, we aim to estimate the $n$ conditional distributions $R_{i}^{*}(w_{i})$ of $Y_{i}$ given $W_{i}=w_{i}$ for $i\in\{1,\ldots,n\}$. To achieve this, we assume (even if this may not hold) that there exists a piecewise constant function $\bsg^{*}$ on $\sW$ such that for all $i\in\{1,\ldots,n\}$, the conditional distributions $R_{i}^{*}(w_{i})$ of $Y_{i}$ given $W_{i}=w_{i}$ are of the form $R_{\bsg^{*}(w_{i})}$ or at least close to it in terms of the Hellinger distance defined in \eref{hellinger-distance-def}. This framework encompasses the following situations:
\begin{enumerate}[label=(\roman*)]
\item The ideal case, where the conditional distributions $R_{i}^{*}(w_{i})=R_{\bsg^{*}(w_{i})}$, for all $i\in\{1,\ldots,n\}$. We then refer to this $\bsg^{*}$ as the {\em regression function} which is a natural generalization from Gaussian regression.
\item The model is slightly misspecified. This includes the situation where for some $i\in\{1,\ldots,n\}$, the conditional distributions $R_{i}^{*}(w_{i})$ are slightly different from the presumed ones $R_{\bsg^{*}(w_{i})}$ or the situation where the data set contains a small amount of outliers.
\end{enumerate}

Let $\sR_{\sW}$ be the collection of all measurable mappings from $(\sW,\cW)$ into $(\sR,\cR)$ and set $\sbR_{\sW}=\sR_{\sW}^{n}$. We denote $\gR^{*}$ the $n$-tuple $(R_{1}^{*},\ldots,R_{n}^{*})$ and hence $\gR^{*}\in\sbR_{\sW}$. Our goal is to estimate this $\gR^{*}$ based on the observations $\bsX=(X_{1},\ldots,X_{n})$. Our estimation strategy is as follows. We suppose that we have at disposal an arbitrary but at most countable collection of (possibly random) piecewise constant candidates of $\bsg^{*}$ mapping $\sW$ into $I$ written as $\widehat\gGamma=\left\{\widehat\bsg_{\lambda}(\bsX),\;\lambda\in\Lambda\right\}$. We then design a procedure based on the same data $\bsX$ to select among $\widehat\gGamma$ and denote the selected one as $\widehat\bsg_{\widehat\lambda}(\bsX)$ (or $\widehat\bsg_{\widehat\lambda}$ for short). Once obtaining $\widehat\bsg_{\widehat\lambda}$, our estimator of $\gR^{*}$ is given by the mapping $\gR_{\widehat\bsg_{\widehat\lambda}}:{\gw}=({w}_{1},\ldots,{w}_{n})\in\sW^{n}\mapsto(R_{\widehat\bsg_{\widehat\lambda}(w_{1})},\ldots,R_{\widehat\bsg_{\widehat\lambda}(w_{n})})$ taking values in $\sQ_{0}^{n}$ with $R_{\widehat\bsg_{\widehat\lambda}}\in\sR_{\sW}$. With a slight abuse of language, sometimes in this paper we also call $\widehat\bsg_{\widehat\lambda}$ an estimator of $\bsg^{*}$ though we know that such a regression function $\bsg^{*}$ does not necessarily exist. It is worth emphasizing that besides the independence, we assume nothing about the distributions of the covariates $W_{i}$ which therefore can be unknown.

To evaluate the performance of the selected estimator $\gR_{\widehat\bsg_{\widehat\lambda}}$, we need to introduce a loss function. Since we focus on estimating the $n$ conditional distributions, it is natural to consider a loss function based on the Hellinger distance. More precisely, we endow the space $\sbR_{\sW}$ with a pseudo Hellinger distance $\gh$ defined for any $\gR=(R_{1},\ldots,R_{n})$ and $\gR'=(R'_{1},\ldots,R'_{n})$ in $\sbR_{\sW}$ by
\begin{align}
\gh^{2}(\gR,\gR')&=\E\cro{\sum_{i=1}^{n}h^{2}\pa{R_{i}(W_{i}),R'_{i}(W_{i})}}=\sum_{i=1}^{n}\int_{\sW}h^{2}\pa{R_{i}(w),R'_{i}(w)}dP_{W_{i}}(w)\label{pseudo-h},
\end{align}
where $h$ is the Hellinger distance introduced in \eref{hellinger-distance-def}. Whenever the regression function $\bsg^{*}$ exists, we automatically deduce a performance of $\widehat\bsg_{\widehat\lambda}$ with respect to $\bsg^{*}$ by the distance $d(\bsg^{*},\widehat\bsg_{\widehat\lambda})=\gh(\gR_{\bsg^{*}},\gR_{\widehat\bsg_{\widehat\lambda}})$. In particular, in the context of changepoint detection problem in exponential families where $W_{i}$ are deterministic, the loss function $\gh$ is the sum of the Hellinger distances between each two probabilities $R_{i}$ and $R'_{i}$. Such a loss function has been considered in several literature, for instance \cite{lecam1986} and \cite{lecamyang1990}. Unlike typical methods that detect changes in distribution parameters (e.g., changes in means for Gaussian or Poisson distributions), our approach identifies changes in the sequence based on abrupt variations at the distribution level.

\section{Non-data-splitting  estimator selection}\label{method}
As already mentioned, given the observations $\bsX=(X_{1},\ldots,X_{n})$, we assume that we have at disposal an arbitrary but at most countable (possibly random) candidates $\widehat\gGamma=\{\widehat \bsg_{\lambda}(\bsX),\; \lambda\in\Lambda\}$ for $\bsg^{*}$, where for each $\lambda\in\Lambda$, $\widehat \bsg_{\lambda}$ is piecewise constant on $\sW$. This $\widehat\gGamma$ may contain the estimators based on the minimization of some criteria, estimators based on Bayes procedures or just simple guesses by some experts. The dependency of these estimators with respect to the observations $\bsX$ can be unknown. Our goal is to select some $\widehat\bsg_{\widehat\lambda}(\bsX)$ among the family $\widehat\gGamma=\{\widehat\bsg_{\lambda}(\bsX),\; \lambda\in\Lambda\}$ based on the same observations $\bsX$ such that the risk of our estimator is as close as possible to the quantity $\inf_{\lambda\in\Lambda}\E[\gh^{2}(\gR^{*},\gR_{\widehat\bsg_{\lambda}})]$.

\subsection{Estimator selection procedure}\label{sele-change-point}
Let $\cM$ be a finite or countable set of partitions on $\sW$. We begin with a family of collections $\left\{\gGamma_{m},\; m\in\cM\right\}$ indexed by the partition $m$ on $\sW$, where for each $m\in\cM$, $\gGamma_{m}$ stands for an at most countable collection of piecewise constant functions on $\sW$ with values in $I$ based on the partition $m$. Setting the notation $\widetilde\gGamma=\cup_{m\in\cM}\gGamma_{m}$, we assume the family of (possibly random) candidates $\widehat\gGamma=\{\widehat \bsg_{\lambda}(\bsX),\; \lambda\in\Lambda\}$ for $\bsg^{*}$ (may not exist) with values in $\widetilde\gGamma$. This is to say, for each $\lambda\in\Lambda$, there is a (possibly random) partition $\widehat m(\lambda)\in\cM$ such that $\widehat\bsg_{\lambda}\in\gGamma_{\widehat m(\lambda)}$.
For any $\bsg\in\widetilde\gGamma$, we define $$\cM(\bsg)=\left\{m\in\cM,\;\bsg\in\gGamma_{m}\right\},$$ therefore naturally we have $\widehat m(\lambda)\in\cM(\widehat\bsg_{\lambda})$.

Let $\Delta(\cdot)$ be a map from $\cM$ to $\R_{+}=[0,+\infty)$. For each $m\in\cM$, we associate it with a nonnegative weight $\Delta(m)$ and assume the following holds true.
\begin{assumption}\label{weight-inequality}
There exists a positive number $\Sigma$ such that
\begin{equation}\label{weight}
\Sigma=\sum_{m\in\cM}e^{-\Delta(m)}<+\infty.
\end{equation}
\end{assumption} 
When $\Sigma=1$, the weights $\Delta(m)$ define a prior distribution on the collection of partitions $\cM$, which gives a Bayesian flavour to our selection procedure. 

Given two partitions $m_{1}, m_{2}\in\cM$, we define a refined partition $m_{1}\vee m_{2}$ on $\sW$ generated by $m_{1},m_{2}$ as $$m_{1}\vee m_{2}=\left\{K_{1}\cap K_{2}\;|\;K_{1}\in m_{1},\;K_{2}\in m_{2},\; K_{1}\cap K_{2}\neq\emptyset\right\}.$$ For any partition $m$ on $\sW$, we denote the number of its segments by $|m|$. To define our selection procedure, we also make the following assumption on the family $\cM$.
\begin{assumption}\label{sum-dim}
There exists some constant $\alpha\geq1$ such that $|m_{1}\vee m_{2}|\leq\alpha(|m_{1}|+|m_{2}|)$, for all $m_{1},m_{2}\in\cM$.
\end{assumption}
Some examples of the family $\cM$ that satisfy Assumption~\ref{sum-dim} are provided below. When $\sW$ is either $\R$ or some subinterval of $\R$, for any finite or countable family $\cM$ of partitions on $\sW$, it is easy to observe that Assumption~\ref{sum-dim} is satisfied with $\alpha=1$. Another example can be the nested partitions, i.e. the family $\cM$ is ordered for the inclusion. In this situation, $m_{1}\vee m_{2}$ either equals to $m_{1}$ or $m_{2}$ so that Assumption~\ref{sum-dim} also holds true with $\alpha=1$. Besides, when $\sW=[0,1)^{d}$ with $d\geq2$, a specific example satisfying Assumption~\ref{sum-dim} with $\alpha=2$ has been introduced in Example~3 of \cite{histogram2006}.

Our selection procedure is based on a pair-by-pair comparison of the candidates, where the selection mechanism is inspired by a series of work of $\rho$-estimation (\cite{MR3595933} and \cite{BarBir2018}). However, unlike the above literature, we generalize the comparison device into the situation where the elements in $\widehat\gGamma$ can be random. 

Let us first introduce a monotone increasing function $\psi$ from $[0,+\infty]$ into $\cro{-1,1}$ defined as
\begin{equation*}
\psi(x)=\left\{
\begin{aligned}
&\frac{x-1}{x+1}&,&\quad\mbox{$x\in[0,+\infty)$,}\\
&1&,&\quad\mbox{$x=+\infty$.}
\end{aligned}
\right.
\end{equation*}
For any $\bsg, \bsg'\in\widetilde\gGamma$, we define the $\gT$-statistic as
\begin{equation*}
\gT(\bsX,\bsg,\bsg')=\sum_{i=1}^{n}\psi\pa{\sqrt{\frac{r_{\bsg'(W_{i})}(Y_{i})}{r_{\bsg(W_{i})}(Y_{i})}}}
\end{equation*}
with the conventions $0/0=1$ and $a/0=+\infty$ for all $a>0$. 
Let $D_{n}$ be a map from $\cM$ to $\R_{+}$ defined as, for any $m\in\cM$, 
\begin{equation*}
D_{n}(m)=|m|\cro{9.11+\log_{+}\left(\frac{n}{|m|}\right)},
\end{equation*}
where $\log_{+}(x)=\max\left\{\log(x),0\right\}$.
We define the penalty function from $\widetilde\gGamma$ to $\R_{+}$ such that for all $\bsg\in\widetilde\gGamma$,
\begin{equation}\label{penalty-procedure}
\mathbf{pen}(\bsg)\geq C_{0}\left(2\alpha+\frac{1}{2}\right)\inf_{m\in\cM(\bsg)}\cro{D_{n}(m)+\Delta(m)},
\end{equation}
where $C_{0}>0$ is a universal constant. For each $\lambda\in\Lambda$, we set
\begin{equation*}
\gup(\bsX,\widehat\bsg_{\lambda})=\sup_{\lambda'\in\Lambda}\cro{\gT(\bsX,\widehat\bsg_{\lambda},\widehat\bsg_{\lambda'})-\gpen(\widehat\bsg_{\lambda'})}+\gpen(\widehat\bsg_{\lambda}). 
\end{equation*}
We select $\widehat\bsg_{\widehat\lambda}$ as any measurable element of the random (and non-void) set
\begin{equation}\label{def-sE-2}
\cE(\bsX)=\ac{\widehat\bsg_{\lambda}\in\widehat\gGamma\;\mbox{ such that }\; \gup(\bsX,\widehat\bsg_{\lambda})\leq \inf_{\lambda'\in\Lambda}\gup(\bsX,\widehat\bsg_{\lambda'})+1}.
\end{equation}
The final selected estimator $\gR_{\widehat\bsg_{\widehat\lambda}}$ of $\gR^{*}$ is given by $\gR_{\widehat\bsg_{\widehat\lambda}}=(R_{\widehat\bsg_{\widehat\lambda}},\ldots,R_{\widehat\bsg_{\widehat\lambda}})$.

The number 1 in \eref{def-sE-2} does not play any role, therefore can be substituted by any small number $\delta>0$. We choose $\delta=1$ here just for enhancing the legibility of our results. Moreover, to improve the performance of the selected estimator $\gR_{\widehat\bsg_{\widehat\lambda}}$, the choice of a $\widehat\bsg_{\widehat\lambda}$ such that $\gup(\bsX,\widehat\bsg_{\widehat\lambda})=\inf_{\lambda\in\Lambda}\gup(\bsX,\widehat\bsg_{\lambda})$ should be preferred whenever available, which is the case when $\widehat\gGamma$ is a finite set.
\subsection{The performance of the selected estimator}\label{estimator-selection}
We establish non-asymptotic exponential inequalities of deviations between the selected estimator $\gR_{\widehat\bsg_{\widehat\lambda}}$ and the truth $\gR^{*}$.
\begin{theorem}\label{thm-1}
Under Assumptions~\ref{weight-inequality} and \ref{sum-dim}, whatever the conditional distributions $\gR^{*}=(R^{*}_{1},\ldots, R^{*}_{n})$ of $Y_{i}$ given $W_{i}$ and the distributions of $W_{i}$, there exists a universal constant $C_{0}>0$ such that the selected estimator $\gR_{\widehat\bsg_{\widehat\lambda}}$ given by the procedure in Section~\ref{sele-change-point} among a family of (possibly random) candidates $\widehat\gGamma=\left\{\widehat\bsg_{\lambda}(\bsX),\; \lambda\in\Lambda\right\}$ based on the observations $\bsX=(X_{1},...,X_{n})$ satisfies for any $\xi>0$, on a set of probability larger than $1-\Sigma^{2}e^{-\xi}$
\begin{equation}\label{performance}
\gh^{2}(\gR^{*},\gR_{\widehat\bsg_{\widehat\lambda}})\leq\inf_{\lambda\in\Lambda}\cro{c_{1}\gh^{2}(\gR^{*},\gR_{\widehat\bsg_{\lambda}})+c_{2}\mathbf{pen}(\widehat\bsg_{\lambda})}+c_{3}\left(1.471+\xi\right),
\end{equation}
where $c_{1}=91.4$, $c_{2}=42.7$ and $c_{3}=12666.9$. 
\end{theorem}
The proof of Theorem~\ref{thm-1} is provided in the supplementary material. We hereby give a short discussion of the numerical constant $C_{0}$ appearing in the penalty function \eref{penalty-procedure}. In the proof of Theorem~\ref{thm-1}, we show that there does exist a numerical constant $C_{0}>0$ such that for all the penalties satisfying \eref{penalty-procedure}, our procedure defined in Section~\ref{sele-change-point} results in a selected estimator fulfilling the performance stated in Theorem~\ref{thm-1}. Unfortunately, this theoretical constant $C_{0}$ turns out to be quite large and we do not have enough information about the smallest value of $C_{0}$ which validates the non-asymptotic exponential inequalities in \eref{performance}. In practice, when we implement our estimator selection procedure we regard this $C_{0}$ as a tuning parameter instead of using the theoretical value. For this point, we will make it more clear in the simulation study, where it also turns out the value of $C_{0}$ in theory seems to be too pessimistic.

To comment on the performance of the selected estimator further, we integrate (\ref{performance}) with respect to $\xi$ and obtain the following risk bound.
\begin{corollary}\label{cor1}
Under Assumptions~\ref{weight-inequality} and \ref{sum-dim}, whatever the conditional distributions $\gR^{*}=(R^{*}_{1},\ldots, R^{*}_{n})$ of $Y_{i}$ given $W_{i}$ and the distributions of $W_{i}$, there exists a universal constant $C_{0}>0$ such that the selected estimator $\gR_{\widehat\bsg_{\widehat\lambda}}$ given by the procedure in Section~\ref{sele-change-point} among $\widehat\gGamma=\left\{\widehat\bsg_{\lambda}(\bsX),\; \lambda\in\Lambda\right\}$ satisfies
\begin{align*}
\E\cro{\gh^{2}(\gR^{*},\gR_{\widehat\bsg_{\widehat\lambda}})}&\leq\E\cro{\inf_{\lambda\in\Lambda}\left(c_{1}\gh^{2}(\gR^{*},\gR_{\widehat\bsg_{\lambda}})+c_{2}\mathbf{pen}(\widehat\bsg_{\lambda})\right)}+c_{3}\left(\Sigma^{2}+1.471\right)\\
&\leq\inf_{\lambda\in\Lambda}\left\{\E\cro{c_{1}\gh^{2}(\gR^{*},\gR_{\widehat\bsg_{\lambda}})+c_{2}\mathbf{pen}(\widehat\bsg_{\lambda})}\right\}+c_{3}\left(\Sigma^{2}+1.471\right).
\end{align*}
In particular, if the equality in \eref{penalty-procedure} holds,
\begin{equation}\label{select-bound}
\E\cro{\gh^{2}(\gR^{*},\gR_{\widehat\bsg_{\widehat\lambda}})}\leq C_{\alpha,\Sigma}\inf_{\lambda\in\Lambda}\left\{\E\cro{\gh^{2}(\gR^{*},\gR_{\widehat\bsg_{\lambda}})}+\E\cro{\bs{\Xi}(\widehat\bsg_{\lambda})}\right\}, 
\end{equation}
where for all $\lambda\in\Lambda$,
\begin{align*}
\bs{\Xi}(\widehat\bsg_{\lambda})&=\inf_{m\in\cM(\widehat\bsg_{\lambda})}\cro{|m|\left(9.11+\log_{+}\left(\frac{n}{|m|}\right)\right)+\Delta(m)}\\
&\leq|\widehat m(\lambda)|\cro{9.11+\log_{+}\left(\frac{n}{|\widehat m(\lambda)|}\right)}+\Delta(\widehat m(\lambda))
\end{align*}
and $$C_{\alpha,\Sigma}=\cro{c_{2}C_{0}\left(2\alpha+\frac{1}{2}\right)+\frac{c_{3}\left(\Sigma^{2}+1.471\right)}{9.11}}\vee c_{1}.$$
\end{corollary}
The result given in \eref{select-bound} compares the risk of the selected estimator $\gR_{\widehat\bsg_{\widehat\lambda}}$ to those of $\gR_{\widehat\bsg_{\lambda}}$ plus an additional nonnegative term $\E\cro{\bs{\Xi}(\widehat\bsg_{\lambda})}$. One nice feature of this approach implied by (\ref{select-bound}) lies in the fact that the risk bound does not depend on the cardinality of the set $\widehat\gGamma$. This entails that if we enlarge the collection of our candidates by keeping $\cM$ unchanged (so that $\Delta(m)$ will not change), the risk bound for the selected estimator only decreases over the larger collection of candidates. On the other hand, our procedure is based on $\cO(|\widehat\gGamma|^{2})$ times of pair-by-pair comparisons. Therefore, the payment for enlarging set $\widehat\gGamma$ is the computation time.

The risk bound \eref{select-bound} in Corollary~\ref{cor1} also accounts for the stability of our selection procedure under a slight misspecification framework. To illustrate, let us first consider the ideal situation where $\gR^{*}=\gR_{\bsg^{*}}=(R_{\bsg^{*}},\ldots,R_{\bsg^{*}})$ with $\bsg^{*}$ a piecewise constant function based on the partition $m^{*}$ of $\sW$. We denote $\bsG_{m^{*}}$ the class of all piecewise constant functions with values in $I\subset\R$ based on the partition $m^{*}$ and assume for simplicity $\widehat\gGamma=\gGamma_{m^{*}}$, where $\gGamma_{m^{*}}$ stands for a dense (for the topology of the pointwise convergence) and countable subset of $\bsG_{m^{*}}$. Taking $\Delta(m^{*})=0$, we deduce from \eref{select-bound} that the estimator $\gR_{\widehat\bsg}$ based on the selection among $\gGamma_{m^{*}}$ satisfies for $C>0$ being a numerical constant,
\begin{equation}\label{rob-single}
\E\cro{\gh^{2}(\gR_{\bsg^{*}},\gR_{\widehat\bsg})}\leq C|m^{*}|\cro{1+\log_{+}\left(\frac{n}{|m^{*}|}\right)},
\end{equation}
which is, up to a logarithm term, the expected magnitude of $|m^{*}|$ for the quantity $\gh^{2}(\gR_{\bsg^{*}},\gR_{\widehat\bsg})$. If it is not the ideal case, an approximation error $\gh^{2}(\gR^{*},\overline\sbQ_{m^{*}})$ with $\overline\sbQ_{m^{*}}=\left\{\gR_{\bsg},\;\bsg\in\bsG_{m^{*}}\right\}$, will be added into the right hand side of \eref{rob-single} according to \eref{select-bound}. However, as long as this bias term remains small, the performance of our selected estimator will not deteriorate too much as compared to the ideal situation.

\subsection{Connection to model selection}
The work done in this paper differs from the corresponding result (12) given by a model selection procedure in \cite{ChenModSel}. In fact, one can regard Corollary~\ref{cor1} in this paper as a more general result of the one in \cite{ChenModSel}. We illustrate this connection as follows.

We consider the particular application of our selection procedure in the context of model selection. For simplicity, let the equality holds in \eref{penalty-procedure}. We take $\Lambda=\{1,\ldots,|\widetilde\gGamma|\}$ which is the index set of all the functions belonging to $\widetilde\gGamma=\cup_{m\in\cM}\gGamma_{m}$ so that in this case, $\widehat\gGamma=\widetilde\gGamma=\left\{\bsg_{\lambda},\;\lambda\in\Lambda\right\}$ is a collection of deterministic candidates. Moreover, for each $\lambda\in\Lambda$, there exists a deterministic $m(\lambda)\in\cM$ such that $\bsg_{\lambda}\in\gGamma_{m(\lambda)}$. Let us denote $\sbQ_{m}=\left\{\gR_{\bsg},\;\bsg\in\gGamma_{m}\right\}$, for all $m\in\cM$. We can immediately deduce from \eref{select-bound} that the estimator $\gR_{\widehat\bsg}$ based on the selection among the family $\left\{\gGamma_{m},\;m\in\cM\right\}$ satisfies
\begin{equation*}
\E\cro{\gh^{2}(\gR^{*},\gR_{\widehat\bsg})}\leq C_{\alpha,\Sigma}\inf_{m\in\cM}\cro{\gh^{2}(\gR^{*},\sbQ_{m})+D_{n}(m)+\Delta(m)},
\end{equation*}
which is, up to constants, the result (12) of \cite{ChenModSel} when one takes $\gGamma_{m}$ in \cite{ChenModSel} as the collection of piecewise constant functions on $\sW$. The difference is their model selection procedure, on the one hand, does not require Assumption~\ref{sum-dim} to be satisfied and can be applied to other types of models to approximate the potential $\bsg^{*}$ besides piecewise constant ones. On the other hand, when the number of models becomes large, model selection strategy is more of theoretical interest due to its expensive numerical cost. Our estimator selection strategy, however, allows to deal with random partitions which can be obtained for example from dynamic programming algorithm (e.g. \cite{rigaill2015pruned}) or CART algorithm (e.g. \cite{breiman1984classification}). Efficiently reducing the cardinality of $\widehat\gGamma$, these algorithms together with our estimator selection procedure take the model selection strategy into practice. Moreover, the idea that selecting among random candidates set makes the selection between estimators given by different model selection strategies possible.

\section{Application to changepoint detection in exponential families}\label{apply-to-changepoint}
In this section, we apply our estimator selection procedure to the changepoint detection problem in exponential families, where the exponential family $\sQ_{0}=\left\{R_{\gamma},\;\gamma\in I\right\}$ is parametrized in its natural form. This entails using the identity function for $u$ in \eref{general-den-2} and $$A(\gamma)=\log\cro{\int_{\sY}\exp\left(\gamma T(y)\right)d\nu(y)}.$$ We observe a sequence $\bsY=(Y_{1},\ldots,Y_{n})$ with values in $\sY^{n}$ and assume that there exists a vector $\gamma^{*}=(\gamma^{*}_{1},\ldots,\gamma^{*}_{n})\in I^{n}$ with $N-1$ changepoints, $N\geq1$ such that within each segment, the values of $\gamma^{*}$ remain a constant and for each $i\in\{1,\ldots,n\}$, the distribution of $Y_{i}$ is given by $R_{\gamma^{*}_{i}}$. This corresponds to the situation in our setting when $W_{i}=(i-1)/n$ are deterministic, for all $i\in\{1,\ldots,n\}$ so that $\sW=[0,1)$ and the function $\bsg^{*}:[0,1)\rightarrow I\subset\R$ is a right-continuous step function with $N\geq1$ segments. For consistency with the previous paragraphs, we take $W_{i}=(i-1)/n$ throughout this section and use the function notation $\bsg^{*}$ rather than the vector $\gamma^{*}\in I^{n}$ in the sequel.

For each $1\leq k\leq n$, let $\cM_{k}$ stand for the collection of all possible partitions of the sequence $1,\ldots,n$ into $k$ segments and denote $\cM=\cup_{1\leq k\leq n}\cM_{k}$. In changepoint detection problem, for each $m\in\cM$, we assign its weight as
\begin{equation}\label{weight-changepoint}
\Delta(m)=\log\binom{n-1}{|m|-1}+|m|.
\end{equation}
With \eref{weight-changepoint}, a basic computation leads to $\Sigma=\sum_{m\in\cM}\exp\cro{-\Delta(m)}\leq1/(e-1)$ which entails Assumption~\ref{weight-inequality} is satisfied. Moreover, since $\sW=[0,1)\subset\R$, for any $m_{1},m_{2}\in\cM$, $|m_{1}\vee m_{2}|\leq|m_{1}|+|m_{2}|-1$, Assumption~\ref{sum-dim} also holds true with $\alpha=1$.

Supposing that we have a finite but arbitrary collection of (possibly random) piecewise constant candidates $\widehat\gGamma=\left\{\widehat\bsg_{\lambda}(\bsX),\;\lambda\in\Lambda\right\}$, we associate each $\widehat\bsg_{\lambda}(\bsX)$ with the penalty
\begin{equation}\label{penalty-changepoint}
\gpen(\widehat\bsg_{\lambda})=\kappa\left\{|\widehat m(\lambda)|\cro{10.11+\log\left(\frac{n}{|\widehat m(\lambda)|}\right)}+\log\binom{n-1}{|\widehat m(\lambda)|-1}\right\},
\end{equation}
where $\kappa$ is the parameter to be tuned later. Once the value of $\kappa$ is given, our estimator selection procedure can be implemented by executing Algorithm~\ref{algo1}.

\begin{algorithm}[h]
\caption{Estimator selection}
\label{algo1}
\begin{algorithmic}[1]
\State \textbf{Input:} $\bsX = (X_1, \cdots, X_n)$: the observations
\State \textbf{Output:} $\widehat{\bsg}_{\widehat{\lambda}}$: the selected estimator
\State Collect $\widehat{\gGamma} = \{\widehat{\bsg}_{\lambda}, \lambda \in \Lambda\}$ based on $\bsX$
\For{$\lambda \in \Lambda$}
    \State $\gup(\bsX, \widehat{\bsg}_{\lambda}) \gets \sup_{\lambda' \in \Lambda} \left(\gT(\bsX, \widehat{\bsg}_{\lambda}, \widehat{\bsg}_{\lambda'}) - \gpen(\widehat{\bsg}_{\lambda'})\right) + \gpen(\widehat{\bsg}_{\lambda})$
\EndFor
\State $\widehat{\lambda} \gets \argmin_{\lambda \in \Lambda} \gup(\bsX, \widehat{\bsg}_{\lambda})$
\State Return $\widehat{\bsg}_{\widehat{\lambda}}$
\end{algorithmic}
\end{algorithm}

\subsection{Calibrating the value of $\kappa$}
We set $\kappa=0.08$ uniformly across all exponential families. The reason for this choice is explained below.

We first simulate data of size $n$ and prepare a collection of candidate estimators, $\widehat\gGamma$, by running the algorithm in the R package \texttt{Segmentor3IsBack} which implements the procedure proposed by \citeauthor{refId0} (\citeyear{refId0}, \citeyear{cleynen2017}). We then choose different values of $\kappa$ to design the penalty function \eref{penalty-changepoint}, yielding a sequence of selected $\widehat\bsg_{\kappa,\widehat\lambda}$ from $\widehat\gGamma$, corresponding to various values of $\kappa$. For each value of $\kappa$, the experiment is repeated 100 times under each simulation setting. The risk $\E[\gh^{2}(\gR^{*},\gR_{\widehat\bsg_{\kappa,\widehat\lambda}})]$ of the selected estimator $\gR_{\widehat\bsg_{\kappa,\widehat\lambda}}$ is then evaluated by its empirical mean:
\[
\widehat R_{n}\left(\widehat\bsg_{\kappa,\widehat\lambda}\right)=\frac{1}{100}\sum_{l=1}^{100}\cro{\sum_{i=1}^{n}h^{2}\pa{R^{*}_{i},R_{\widehat\bsg^{l}_{\kappa,\widehat\lambda}\left(\frac{i-1}{n}\right)}}},
\]
where $\widehat\bsg^{l}_{\kappa,\widehat\lambda}$ is the $l$-th realization of the selected estimator associated with a fixed $\kappa$. 

\subsubsection{Data simulation}\label{sec-4.1.1}
We conduct experiments on three models: Gaussian, Poisson, and exponential changepoint detection.

Let $\bsg^{*}$ be piecewise constant on $[0,1)$ with $N$ segments and $\gR^{*}=\gR_{\bsg^{*}}$. For each model, we design the experiments under three settings where for all the settings $n=500$, but $N=5$, $N=10$ and $N=20$ respectively. In all three settings, the changepoints are uniformly spaced: every 100 data points in the first setting, every 50 data points in the second, and every 25 data points in the third.

\begin{figure}
        \begin{subfigure}{\linewidth}
\includegraphics[width=\linewidth]{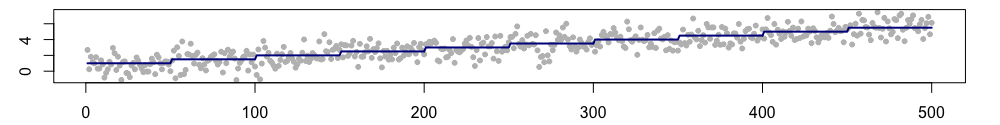}
    \end{subfigure}
    \begin{subfigure}{\linewidth}
\includegraphics[width=\linewidth]{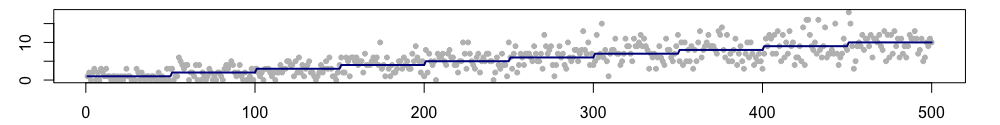}
    \end{subfigure}
    \begin{subfigure}{\linewidth}
\includegraphics[width=\linewidth]{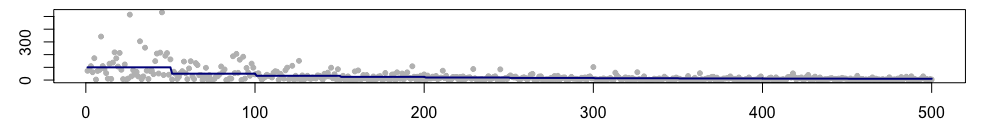}
    \end{subfigure}
    \caption{The 1st graph (top) corresponds to one profile of the simulated data (dots) and $\bsg^{*}$ (solid line) for Gaussian model; The 2nd graph (middle) corresponds to one profile of the simulated data (dots) and $\exp(\bsg^{*})$ (solid line) for Poisson model; The 3rd graph (bottom) corresponds to one profile of the simulated data (dots) and $1/\bsg^{*}$ (solid line) for exponential model.}
    \label{fig-exm-cal}
\end{figure}

\begin{enumerate}[label=(\roman*)]
\item In all the settings of Gaussian model, for $1\leq i\leq n$, if $Y_{i}$ is in the $j$-th segment with $1\leq j\leq N$, $Y_{i}$ follows a Gaussian distribution with mean $(j+1)/2$, variance $\sigma^{2}=1$. 
\item In all the settings of Poisson model, for $1\leq i\leq n$, if $Y_{i}$ is in the $j$-th segment with $1\leq j\leq N$, $Y_{i}$ follows a Poisson distribution with mean $j$ which means $\bsg^{*}$ takes value $\log(j)$ on the $j$-th segment. 
\item In all the settings of exponential model, for $1\leq i\leq n$, if $Y_{i}$ is in the $j$-th segment with $1\leq j\leq N$, $Y_{i}$ follows an exponential distribution with natural parameter $0.01j$. 

Figure~\ref{fig-exm-cal} exhibits one example of the simulated data (when $N=10$) and the true value of the regression function $\bsg^{*}$ (or a suitable transformation of $\bsg^{*}$) on each segment. 
\end{enumerate}

\subsubsection{Gathering candidates for the random set}
In \citeauthor{refId0} (\citeyear{refId0}, \citeyear{cleynen2017}), the problem was solved through a model selection procedure using a penalty function, based on partitions from the pruned dynamic programming algorithm (PDPA) proposed by \cite{rigaill2015pruned}. Given $N_{\max}$ as the maximum number of segments for consideration, for each integer $\lambda$ with $1\leq\lambda\leq N_{\max}$, PDPA searches for the optimal partition with exactly $\lambda$ segments. With $N_{\max}=30$, PDPA returns 30 different partitions of the sequence $1,\ldots,n$. For each partition, the value of $\bsg^{*}$ on each segment is estimated using maximum likelihood, as done in \citeauthor{refId0} (\citeyear{refId0}, \citeyear{cleynen2017}). This results in 30 candidates, denoted as $\widehat\gGamma_{c}=\{\widehat\bsg_{\lambda}, 1\leq\lambda\leq N_{\max}\}$.

\subsubsection{Results}
\begin{figure}[t]
    \centering
    \begin{subfigure}[t]{0.45\textwidth}
        \centering
        \includegraphics[width=\textwidth,height=3.5cm]{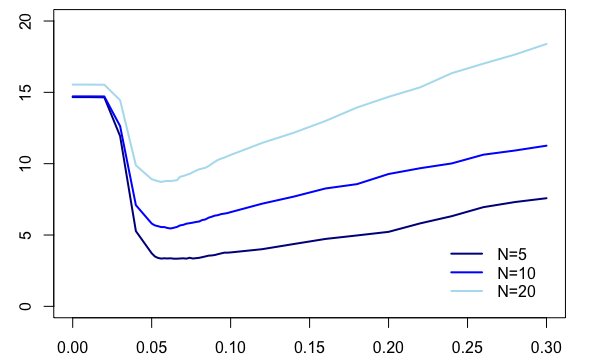}
\caption{Gaussian}
    \end{subfigure}
    \hfill
    \begin{subfigure}[t]{0.45\textwidth}
        \centering
        \includegraphics[width=\textwidth,height=3.5cm]{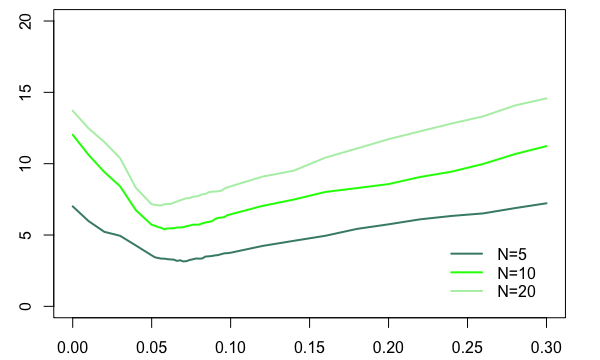}
\caption{Poisson}
    \end{subfigure}
\vskip\baselineskip
    \begin{subfigure}[t]{0.45\textwidth}
        \centering
\includegraphics[width=\textwidth,height=3.5cm]{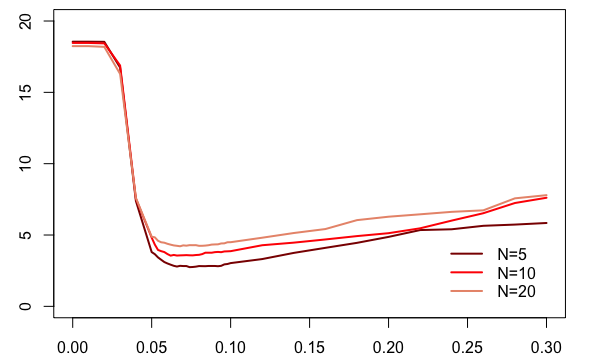}
\caption{exponential}
    \end{subfigure}
    \hfill
    \begin{subfigure}[t]{0.45\textwidth}
        \centering
\includegraphics[width=\textwidth,height=3.5cm]{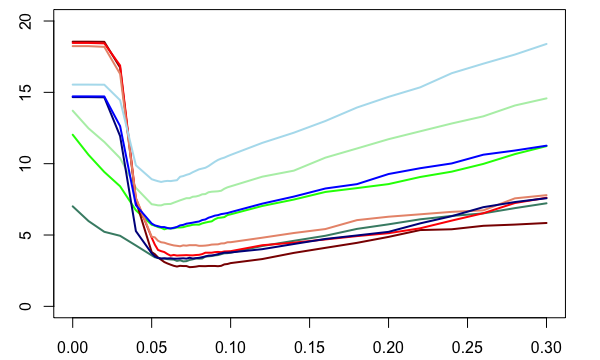}
\caption{3 in 1}
    \end{subfigure}
\caption{$\widehat R_{n}(\widehat\bsg_{\kappa,\widehat\lambda})$ with respect to $\kappa$ under nine settings.}
\label{figR-c}
\end{figure}

In each setting of all the three models, one experiment means we simulate $n=500$ observations with $N$ segments based on the corresponding $\bsg^{*}$ introduced in Section~\ref{sec-4.1.1}. We then select the estimator from the candidates $\widehat\gGamma_{c}$ using penalty functions \eqref{penalty-changepoint} corresponding to different values of $\kappa$. Finally, we record the quantity $\widehat R_{n}(\widehat\bsg_{\kappa,\widehat\lambda})$ and use it as the criterion to calibrate an appropriate value of $\kappa$. The results for all nine settings are shown in Figure~\ref{figR-c}, where the horizontal axis represents the value of $\kappa$ and the vertical axis represents the quantity $\widehat R_{n}(\widehat\bsg_{\kappa,\widehat\lambda})$.

In Figure~\ref{figR-c}, the quantities $\widehat R_{n}(\widehat\bsg_{\kappa,\widehat\lambda})$ for all nine settings initially decrease and then increase as $\kappa$ increases, which aligns with the theoretical results. When $\kappa$ is too small, the penalty function is relatively weak for more complex models, leading to potential overfitting. Conversely, when $\kappa$ is large, the penalty function becomes excessively strong, resulting in overpenalization. Moreover, the minimizers of $\widehat R_{n}(\widehat\bsg_{\kappa,\widehat\lambda})$ across all nine settings are closely clustered within the interval $\cro{0.05,0.1}$. Considering the optimal performance across all settings and aiming to avoid overfitting, we select $\kappa=0.08$, the largest minimizer, for our procedure in subsequent studies.

\section{Simulation study and discussion}\label{simulation}
Throughout this section, we conduct a comparative simulation study with state-of-the-art competitors available in R packages for the changepoint detection problem in exponential families. Unless otherwise specified, the competitors are implemented using the default settings in their respective packages. For the Gaussian model, some of our competitors use the estimated value of the standard deviation $\sigma$. To ensure a fair comparison, we also implement the median absolute deviation estimator for $\sigma$ while running our procedure, as adopted in \cite{Killick_2012} and \cite{ChaDetectOut}.

To evaluate the performance of each estimator, besides the empirical risk $\widehat R_{n}(\cdot)$ obtained from replications, we also record $\widehat N-N$ which computes the difference between the estimated number of segments and the truth for each replication.

\subsection{Accuracy}\label{accuracy}
In this section, we study the changepoint detection problem for the Gaussian model, a topic extensively explored in the literature. We construct our candidate set $\widehat\gGamma$ as a collection of cutting-edge estimators implemented in R packages, which also serve as competitors to our estimator, ES. Specifically, the competing packages considered are: \texttt{PSCBS}, implementing the CBS procedure from \cite{PMID:15475419}; \texttt{cumSeg}, based on the method in \cite{10.1093/bioinformatics/btq647}; \texttt{changepoint}, implementing the PELT approach from \cite{Killick_2012}; \texttt{StepR}, implementing SMUCE from \cite{Frick2013MultiscaleCP}; \texttt{Segmentor3IsBack}, implementing CL proposed by \citeauthor{refId0} (\citeyear{refId0}, \citeyear{cleynen2017}); \texttt{wbs}, implementing wild binary segmentation from \cite{Fryzlewicz2014}; \texttt{FDRSeg}, based on the approach in \cite{2016fdr}; and \texttt{robseg}, implementing the procedure from \cite{ChaDetectOut}. We aim to evaluate the performance of our estimator, ES, through comparison with these state-of-the-art methods.\vspace{10pt}

\begin{figure}[h]
     \begin{subfigure}{0.32\textwidth}
    \includegraphics[width=\textwidth,height=3cm]{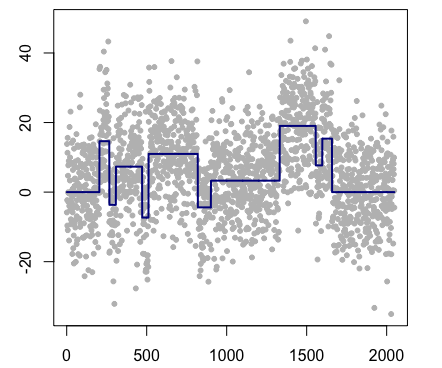}
         \caption{blocks} 
     \end{subfigure}
     \hfill
     \begin{subfigure}{0.32\textwidth}
         \includegraphics[width=\textwidth,height=3cm]{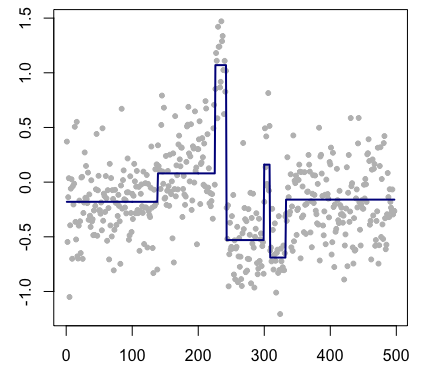}
         \caption{fms ($\sigma=0.3$)} 
     \end{subfigure}
     \hfill
      \begin{subfigure}{0.32\textwidth}
         \includegraphics[width=\textwidth,height=3cm]{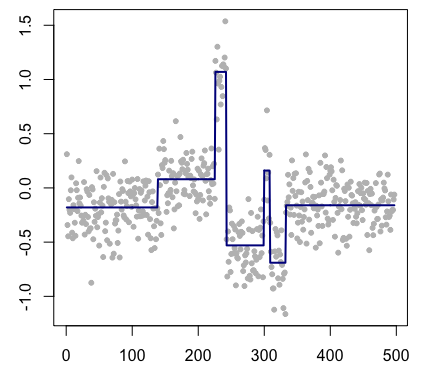}
         \caption{fms ($\sigma=0.2$)} 
     \end{subfigure}
\vskip\baselineskip
     \begin{subfigure}{0.32\textwidth}
    \includegraphics[width=\textwidth,height=3cm]{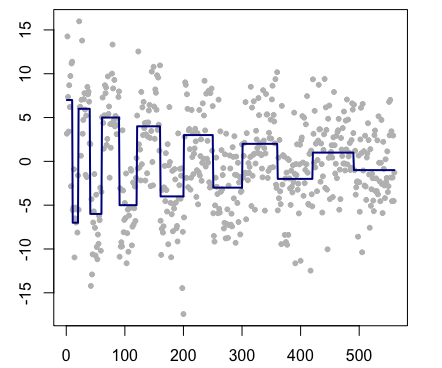}
         \caption{mix} 
     \end{subfigure}
     \hfill
     \begin{subfigure}{0.32\textwidth}
         \centering
         \includegraphics[width=\textwidth,height=3cm]{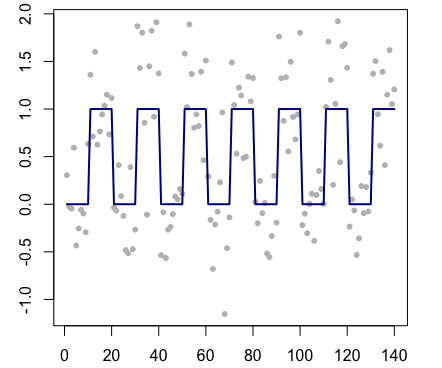}
         \caption{teeth10} 
     \end{subfigure}
     \hfill
      \begin{subfigure}{0.32\textwidth}
         \centering
         \includegraphics[width=\textwidth,height=3cm]{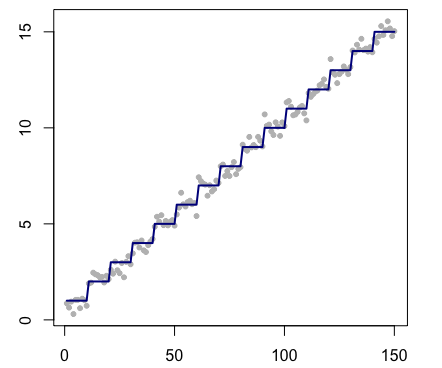}
         \caption{stairs10} 
     \end{subfigure}
        \caption{The six signals (solid line) and simulated data (dots).} 
        \label{six-signal}
\end{figure}

We follow the test signals used by \cite{Fryzlewicz2014} and later by \cite{ChaDetectOut}, which include five different signal formats with lengths ranging from $n=140$ to 2048: (1) \texttt{blocks}, (2) \texttt{fms}, (3) \texttt{mix}, (4) \texttt{teeth10}, and (5) \texttt{stairs10}. The specific settings of these signals, including sample sizes and noise standard deviations, are provided in Appendix B of \cite{Fryzlewicz2014}. Following the experiments in \cite{ChaDetectOut}, we also consider an additional setting by changing the standard deviation of (2) \texttt{fms} from 0.3 to 0.2, as studied in \cite{Frick2013MultiscaleCP}. An example of a simulated data profile along with the underlying signals $\bsg^{*}$ is shown in Figure~\ref{six-signal}. For each signal, the experiment was replicated 1000 times. The results are presented in Table~\ref{result-1}, and the performance of each estimator is described as follows.

CBS and cumSeg. The CBS and cumSeg in general behave poorly compared with other procedures. The CBS only has satisfactory performance of detecting changes for \texttt{blocks} and \texttt{fms} ($\sigma=0.2$) but it turns out CBS always results in a relatively large empirical risk $\widehat R_{n}(\cdot)$. Except acceptable performance for \texttt{fms} ($\sigma=0.2$) and \texttt{stairs10}, cumSeg always tends to underestimate the number of changes and also yields an estimator with quite large empirical risk.

PELT. The PELT has excellent performance for both of the \texttt{fms} signals and \texttt{stairs10}. For \texttt{blocks} signal, it is above the average but does not belong to the first class among all. As for \texttt{mix} and \texttt{teeth10}, it performs rather average.

SMUCE. The SMUCE has very excellent performance for \texttt{fms} ($\sigma=0.2$). However, it behaves poorly for all the other signals.

CL. The CL has nice performance for \texttt{teeth10}. For \texttt{blocks} and \texttt{mix}, its performance is satisfactory though not belonging to the first class. For both of the \texttt{fms} signals, it shows rather average performance. The CL does not behave well for the \texttt{stairs10} signal where it tends to overestimate the number of changes compared to other methods. 

WBS sSIC. We implement the package \texttt{wbs} combining the WBS method with the sSIC stopping criterion which, as it has been shown in \cite{Fryzlewicz2014}, is the overall winner compared to combining the WBS method with other thresholding stopping rules. The WBS sSIC has excellent performance for both of the \texttt{fms} signals and \texttt{teeth10}. However, it performs rather average for \texttt{blocks} and \texttt{mix}. As for \texttt{stairs10}, the performance of WBS sSIC is a little poor as a consequence of overestimating the number of changepoints. Such a result has also been confirmed by the study of WBS sSIC in \cite{Fryzlewicz2014}.

FDR. The FDR with $\alpha=0.05$ performs well for \texttt{fms} ($\sigma=0.3$) and \texttt{satirs10} signals. For \texttt{fms} ($\sigma=0.2$), it has an average performance. But it behaves below the average under other test signals.

robseg. We consider Huber loss and biweight loss when implementing the package \texttt{robseg} which are the recommended ones (especially the biweight loss) according to \cite{ChaDetectOut}. The parameter settings follow the suggested values from Section 5.2 of their paper. \texttt{robseg} (Huber) performs excellently on \texttt{blocks}, \texttt{mix}, and \texttt{teeth10}, but shows average performance on both \texttt{fms} signals and \texttt{stairs10}. In contrast, \texttt{robseg} (biweight) excels on both \texttt{fms} signals and \texttt{stairs10}, while its performance on \texttt{blocks}, \texttt{mix}, and \texttt{teeth10} is relatively average.

ES. As we can observe from the column named ``Contribution'' in Table~\ref{result-1}, under different test signals, our estimator selection procedure tends to allocate different preference to the candidates in $\widehat\gGamma$ based on their practical performance. For example, when SMUCE shows obvious outperformance for the signal \texttt{fms} ($\sigma=0.2$), we select it with a frequency 0.734 as our ES estimator. However, when SMUCE performs poorly under other signals we automatically 

{\small\renewcommand*{\arraystretch}{0.43}
\setlength\tabcolsep{2pt}
\begin{longtable}{cc|cc>{\columncolor[gray]{0.8}}ccc|>{\columncolor[gray]{0.8}}c|c}
\caption{Frequencies of $\widehat N-N$ and $\widehat R_{n}(\cdot)$ of ES and its competitors for Gaussian model over 1000 simulated sample paths. Contribution denotes the frequency of each competitor being selected as ES. Bold: highest empirical frequency of $\widehat N-N=0$ and those with frequencies within 10\% off the highest. The uncertainty is obtained by computing $2\widehat\sigma/\sqrt{n_{r}}$, where $\widehat\sigma^{2}$ is the empirical variance and $n_{r}$ is the number of replications.}\vspace{20pt}
\label{result-1}\\
\hline\hline
& &\multicolumn{5}{c|}{$\widehat N-N$} & &\\
\cline{3-7}
Method& Signal& $\leq-2$& -1& 0& 1& $\geq2$& $\widehat R_{n}(\cdot)$&Contribution\\
\hline
ES&blocks& 0.005& 0.278& \bf{0.656}& 0.055& 0.006&5.61\ $\pm$\ 0.12&-\\
CBS&blocks& 0.006& 0.090& 0.575& 0.184& 0.145& 7.57\ $\pm$\ 0.14&0.000\\
cumSeg&blocks& 0.653& 0.335& 0.011& 0.001& 0.000&15.71\ $\pm$\ 0.40&0.000\\
PELT&blocks& 0.014& 0.389& 0.574& 0.020& 0.003&5.69\ $\pm$\ 0.11&0.035\\
SMUCE&blocks& 0.940& 0.060&0.000& 0.000& 0.000&16.02\ $\pm$\ 0.37&0.010\\
CL&blocks& 0.010& 0.356& 0.595& 0.035& 0.004&5.67\ $\pm$\ 0.12&0.533\\
WBS sSIC&blocks& 0.021& 0.412& 0.532& 0.032& 0.003&6.11\ $\pm$\ 0.13&0.013\\
FDR($\alpha=0.05$)&blocks& 0.008& 0.447& 0.478& 0.059& 0.008&6.15\ $\pm$\ 0.13&0.332\\
robseg(Huber)&blocks& 0.004& 0.234& \bf{0.674}& 0.072& 0.016&5.84\ $\pm$\ 0.12&0.063\\
robseg(biweight)&blocks& 0.020& 0.404&0.558& 0.017& 0.001&5.88\ $\pm$\ 0.12&0.014\\
\hline
ES&fms(0.3)& 0.008& 0.002&\bf{0.915}& 0.069& 0.006&2.16\ $\pm$\ 0.07&-\\
CBS&fms(0.3)& 0.007& 0.012& 0.796& 0.139& 0.046& 5.10\ $\pm$\ 0.09&0.000\\
cumSeg&fms(0.3)& 0.706& 0.041& 0.224& 0.028& 0.001&7.07\ $\pm$\ 0.44&0.000\\
PELT&fms(0.3)& 0.007& 0.003& \bf{0.922}& 0.061& 0.007&2.15\ $\pm$\ 0.08&0.054\\
SMUCE&fms(0.3)& 0.074& 0.537&0.388& 0.001& 0.000&5.15\ $\pm$\ 0.18&0.293\\
CL&fms(0.3)& 0.002& 0.001& 0.837& 0.119& 0.041&2.28\ $\pm$\ 0.08&0.199\\
WBS sSIC&fms(0.3)& 0.007& 0.003& \bf{0.933}& 0.048& 0.009&2.26\ $\pm$\ 0.08&0.008\\
FDR($\alpha=0.05$)&fms(0.3)& 0.001& 0.027& \bf{0.879}& 0.076& 0.017&2.28\ $\pm$\ 0.09&0.409\\
robseg(Huber)&fms(0.3)& 0.001& 0.001&0.825& 0.130& 0.043&2.37\ $\pm$\ 0.08&0.007\\
robseg(biweight)&fms(0.3)& 0.013& 0.005&\bf{0.928}& 0.049& 0.005&2.23\ $\pm$\ 0.08&0.030\\
\hline
ES&fms(0.2)& 0.000& 0.000&\bf{0.923}& 0.071& 0.006&1.61\ $\pm$\ 0.06&-\\
CBS&fms(0.2)& 0.000& 0.000&0.871& 0.086& 0.043& 5.79\ $\pm$\ 0.07&0.000\\
cumSeg&fms(0.2)& 0.094& 0.009& 0.812& 0.083& 0.002&5.19\ $\pm$\ 0.22&0.002\\
PELT&fms(0.2)& 0.000& 0.000& \bf{0.929}& 0.060& 0.011&1.59\ $\pm$\ 0.06&0.022\\
SMUCE&fms(0.2)& 0.000& 0.001&\bf{0.994}& 0.005& 0.000&1.49\ $\pm$\ 0.06&0.734\\
CL&fms(0.2)& 0.000& 0.000& 0.840& 0.128& 0.032&1.74\ $\pm$\ 0.07&0.102\\
WBS sSIC&fms(0.2)& 0.000& 0.000& \bf{0.945}& 0.050& 0.005&1.65\ $\pm$\ 0.06&0.003\\
FDR($\alpha=0.05$)&fms(0.2)& 0.000& 0.000& 0.871& 0.103& 0.026&1.66\ $\pm$\ 0.06&0.115\\
robseg(Huber)&fms(0.2)& 0.000& 0.000&0.830& 0.135& 0.035&1.83\ $\pm$\ 0.07&0.008\\
robseg(biweight)&fms(0.2)& 0.000& 0.000&\bf{0.937}& 0.058& 0.005&1.63\ $\pm$\ 0.06&0.014\\
\hline
ES&mix& 0.264& 0.243&\bf{0.434}& 0.056& 0.003&5.91\ $\pm$\ 0.12&-\\
CBS&mix& 0.313& 0.201& 0.324& 0.109& 0.053&11.18\ $\pm$\ 0.17&0.000\\
cumSeg&mix& 0.999& 0.001& 0.000& 0.000& 0.000&32.61\ $\pm$\ 0.92&0.000\\
PELT&mix& 0.375& 0.270&0.321& 0.032& 0.002&6.11\ $\pm$\ 0.12&0.070\\
SMUCE&mix& 0.922& 0.076&0.002& 0.000& 0.000&12.59\ $\pm$\ 0.42&0.042\\
CL&mix& 0.305& 0.244& 0.390& 0.053& 0.008&6.04\ $\pm$\ 0.12&0.585\\
WBS sSIC&mix& 0.342& 0.269& 0.351& 0.032& 0.006&5.99\ $\pm$\ 0.12&0.029\\
FDR($\alpha=0.05$)&mix& 0.411& 0.358& 0.181& 0.038& 0.012&6.71\ $\pm$\ 0.13&0.190\\
robseg(Huber)&mix& 0.209& 0.240&\bf{0.444}& 0.088& 0.019&6.10\ $\pm$\ 0.12&0.051\\
robseg(biweight)&mix& 0.403& 0.264&0.305& 0.026& 0.002&6.30\ $\pm$\ 0.12&0.033\\
\hline
\endfirsthead
\hline\hline
Method& Signal& $\leq-2$& -1& 0& 1& $\geq2$& $\widehat R_{n}(\cdot)$&Contribution\\
\hline
ES&teeth10& 0.215& 0.025&\bf{0.721}& 0.037& 0.002&5.69\ $\pm$\ 0.24&-\\
CBS&teeth10& 0.999& 0.000& 0.001& 0.000& 0.000&24.69\ $\pm$\ 0.07&0.000\\
cumSeg&teeth10& 1.000& 0.000& 0.000& 0.000& 0.000&24.85\ $\pm$\ 0.01&0.005\\
PELT&teeth10& 0.274& 0.029&0.657& 0.037& 0.003&6.03\ $\pm$\ 0.24&0.090\\
SMUCE&teeth10& 0.984& 0.013&0.003& 0.000& 0.000&20.11\ $\pm$\ 0.22&0.003\\
CL&teeth10& 0.029& 0.013&\bf{0.679}& 0.204& 0.075&4.71\ $\pm$\ 0.13&0.321\\
WBS sSIC&teeth10& 0.067& 0.021&\bf{0.752}& 0.120& 0.040&5.30\ $\pm$\ 0.26&0.010\\
FDR($\alpha=0.05$)&teeth10& 0.309& 0.135& 0.508& 0.040& 0.008&7.68\ $\pm$\ 0.32&0.356\\
robseg(Huber)&teeth10& 0.105& 0.026&\bf{0.748}& 0.102& 0.019&4.94\ $\pm$\ 0.15&0.016\\
robseg(biweight)&teeth10& 0.318& 0.028&0.635& 0.019& 0.000&6.31\ $\pm$\ 0.25&0.199\\
\hline
ES&stairs10& 0.00& 0.004&\bf{0.949}& 0.044& 0.003&3.33\ $\pm$\ 0.09&-\\
CBS&stairs10& 0.012& 0.172& 0.789& 0.027& 0.000&13.81\ $\pm$\ 0.16&0.000\\
cumSeg&stairs10& 0.024& 0.090& 0.819& 0.067& 0.000&8.61\ $\pm$\ 0.24&0.000\\
PELT&stairs10& 0.000& 0.004&\bf{0.955}& 0.039& 0.002&3.32\ $\pm$\ 0.09&0.017\\
SMUCE&stairs10& 0.801& 0.137&0.062& 0.000& 0.000&22.26\ $\pm$\ 0.58&0.050\\
CL&stairs10& 0.000& 0.001& 0.768& 0.184& 0.047&3.50\ $\pm$\ 0.09&0.178\\
WBS sSIC&stairs10& 0.000& 0.001& 0.608& 0.301& 0.090&3.91\ $\pm$\ 0.10&0.004\\
FDR($\alpha=0.05$)&stairs10& 0.002& 0.028&\bf{0.896}& 0.053& 0.021&3.57\ $\pm$\ 0.12&0.703\\
robseg(Huber)&stairs10& 0.000& 0.000&0.867& 0.110& 0.023&3.45\ $\pm$\ 0.09&0.006\\
robseg(biweight)&stairs10& 0.000& 0.005&\bf{0.964}& 0.031& 0.000&3.36\ $\pm$\ 0.09&0.042\\
\hline\hline
\endlastfoot
\end{longtable}}
\noindent reduce the frequency to select it as ES but prefer some more competitive ones. As a final result, our ES estimator shows a very competitive performance under all the test signals. The interesting point is that this cannot be achieved by any single candidate in $\widehat\gGamma$ since as we have seen above, each of them only outperforms others for some of the test signals but not all. 

\subsection{Stability when outliers present}\label{robust}
As noted in the theoretical analysis, our estimator selection procedure remains stable even with slight deviations from the assumption $\gR^{*}=\gR_{\bsg^{*}}$, where $\bsg^{*}$ is piecewise constant on $\sW$. This property is particularly relevant when a small proportion of outliers is present in the data, a scenario that has received increasing attention in changepoint detection. In this section, we evaluate the practical performance of ES and its competitors in the presence of outliers. Using the \texttt{fms} signal ($\sigma=0.2$), which is known to perform well with existing methods, we introduce outliers by randomly selecting five points from a sequence of length $n=497$ and altering their values to 3. The results of all estimators are shown in Table~\ref{result-2}.

We can observe from Table~\ref{result-2} that in such a scenario PELT, SMUCE, CL, WBS sSIC, FDR and robseg (Huber) are all not robust with respect to the outliers and they all overestimate the number of changepoints due to fitting the outliers. The CBS and cumSeg still systematically underestimate the number of changepoints. It is not that surprising robseg (biweight) proposed in \cite{ChaDetectOut} is quite robust in this scenario since it was designed to handle such an issue. It shows a very high frequency 0.956 to recover the correct number of changepoints. Moreover, from the quantity of empirical risk $\widehat R_{n}(\cdot)$, it turns out robseg (biweight) outperforms all the other candidates significantly which also indicates an excellent performance of localising the changepoints as well as estimating the value of $\bsg^{*}$ on each segment. Our selection procedure automatically gives the preference to robseg (biweight) in this case with frequency 1.000 which confirms the stability of our selection rule practically. 
\begin{center}
{\small\renewcommand*{\arraystretch}{0.43}
\setlength\tabcolsep{2pt}
\begin{longtable}{ccc|cc>{\columncolor[gray]{0.8}}ccc|>{\columncolor[gray]{0.8}}c|c}
\caption{Frequencies of $\widehat N-N$ and $\widehat R_{n}(\cdot)$ of ES and its competitors for \texttt{fms} ($\sigma=0.2$) signal with 5 outliers over 1000 simulated sample paths.}\vspace{10pt}
\label{result-2}\\
\hline\hline
 & & &\multicolumn{5}{c|}{$\widehat N-N$}&&\\
 \cline{4-8}
Method& Signal&Outlier& $\leq-2$& -1& 0& 1& $\geq2$& $\widehat R_{n}(\cdot)$&Contribution\\
\hline
ES&fms(0.2)&Yes& 0.000& 0.000& \bf{0.956}& 0.043& 0.001&1.64\ $\pm$\ 0.06&-\\
CBS&fms(0.2)&Yes&0.660& 0.282& 0.038& 0.016& 0.004&34.55\ $\pm$\ 0.79&0.000\\
cumSeg&fms(0.2)&Yes& 0.801& 0.056& 0.083& 0.021& 0.039&16.96\ $\pm$\ 0.51&0.000\\
PELT&fms(0.2)&Yes&0.000& 0.000& 0.000& 0.000& 1.000&7.27\ $\pm$\ 0.07&0.000\\
SMUCE&fms(0.2)&Yes&0.000& 0.000& 0.000& 0.000& 1.000&8.02\ $\pm$\ 0.11&0.000\\
CL&fms(0.2)&Yes&0.000& 0.000& 0.000& 0.000& 1.000&7.29\ $\pm$\ 0.07&0.000\\
WBS sSIC&fms(0.2)&Yes&0.000& 0.000& 0.000& 0.000& 1.000&7.33\ $\pm$\ 0.07&0.000\\
FDR($\alpha=0.05$)&fms(0.2)&Yes&0.000& 0.000& 0.000& 0.000& 1.000&7.44\ $\pm$\ 0.07&0.000\\
robseg(Huber)&fms(0.2)&Yes&0.000& 0.000& 0.000& 0.000& 1.000&7.51\ $\pm$\ 0.08&0.000\\
robseg(biweight)&fms(0.2)&Yes& 0.000& 0.000& \bf{0.956}& 0.043& 0.001&1.64\ $\pm$\ 0.06&1.000\\
\hline\hline
\endfirsthead
\end{longtable}
}
\end{center}

\subsection{From Gaussian to Poisson and exponential models}\label{poi-exp-sim}
As mentioned in the introduction, there are few works in the statistical literature addressing changepoint detection for Poisson and exponential models with theoretical guarantees for the proposed estimators. The CL method proposed by \citeauthor{refId0} (\citeyear{refId0}, \citeyear{cleynen2017}) performs model selection based on partitions from \cite{rigaill2015pruned} and establishes that the resulting estimator satisfies an oracle inequality. Their procedure is implemented in the R package \texttt{Segmentor3IsBack}, which handles both Poisson and exponential models. Another approach, proposed by \cite{Frick2013MultiscaleCP} and implemented in the R package \texttt{StepR}, is available only for Poisson segmentation.

Relying solely on these two estimators would be limiting. Recall from Section~\ref{estimator-selection} that a key feature of our selection procedure is the ability to enlarge the (possibly random) collection $\widehat\gGamma$ while keeping $\cM$ unchanged. This ensures that the risk bound for the selected estimator decreases—or at least remains unchanged—when expanding the collection. Therefore, for Poisson and exponential models, in addition to CL and SMUCE (where applicable), we incorporate other reasonable estimators into our candidate set $\widehat\gGamma$. Although these estimators are not established in the literature and lack quantitative analysis, if selected as ES by our procedure, the theoretical guarantee from Section~\ref{estimator-selection} implies that, up to a constant, they outperform the state-of-the-art methods (CL and SMUCE).

A natural idea is to leverage estimators from the Gaussian model, which has been extensively studied. Inspired by \cite{brown2010}, where a mean-matching variance stabilizing transformation (MM-VST) was applied to convert regression problems in exponential families into standard homoscedastic Gaussian regression, we adopt a similar technique for the observations $\bsY$. For details on MM-VST, we refer to Section 2 of \cite{brown2010}. It is worth noting that implementing MM-VST requires selecting the value of $m$, which determines the number of data points binned for the transformation. Although $m$ needs to be suitably chosen for regression problems (see Section 4 of \cite{brown2010}), we avoid using any segmentation-related information in this preprocessing step, as we are dealing with changepoint detection. Therefore, we simply set $m=1$ in the transformation procedure, applying $Y'_{i}=2\sqrt{Y_{i}+1/4}$ for the Poisson model and $Y'_{i}=\log(2Y_{i})$ for the exponential model to obtain new observation sequences $\bsY'=(Y'_{1},\ldots,Y'_{n})$. We then apply the algorithms introduced in the previous section to $\bsY'$ to identify changepoint locations. Based on these locations, we use the $\rho$-estimators proposed in \cite{Baraud2020} to estimate $\bsg^{*}$ on each segment, enhancing performance. As shown in \cite{Baraud2020}, under suitable conditions and when the model is correctly specified, the $\rho$-estimator matches the accuracy of the MLE. Moreover, it offers greater robustness than the MLE in cases of model misspecification or data contamination.

To conclude, the candidates set for the Poisson model is given by
\begin{align}
\widehat\gGamma=&\left\{\SMUCE, \CCL, \mbox{CBS}_{t}+\rho, \mbox{cumSeg}_{t}+\rho, \mbox{PELT}_{t}+\rho, \mbox{WBS sSIC}_{t}+\rho,\right.\nonumber\\
&\ \ \left. \mbox{FDR}_{t}(\alpha=0.05)+\rho, \mbox{robseg(Huber)}_{t}+\rho, \mbox{robseg(biweight)}_{t}+\rho\right\},\label{poi-gam}
\end{align}
that the procedure is implemented on the transformed data. For the exponential model, $\widehat\gGamma$ is constructed similarly to \eref{poi-gam}, except that SMUCE is replaced by $\mbox{SMUCE}_{t}+\rho$ since it is no longer available.

\begin{figure}[t]
     \begin{subfigure}[b]{0.24\textwidth}
         \includegraphics[width=\textwidth]{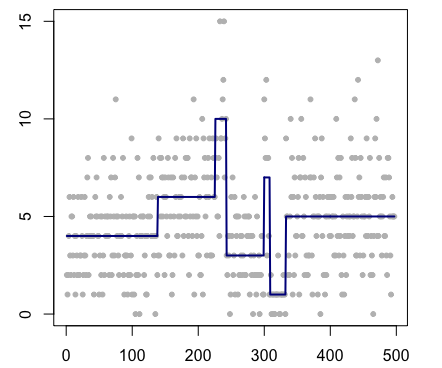}
         \caption{fms-type} 
     \end{subfigure}
     \begin{subfigure}[b]{0.24\textwidth}
         \includegraphics[width=\textwidth]{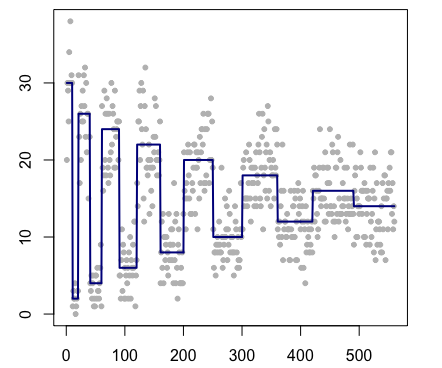}
         \caption{mix-type} 
     \end{subfigure}
                \begin{subfigure}[b]{0.24\textwidth}
         \includegraphics[width=\textwidth]{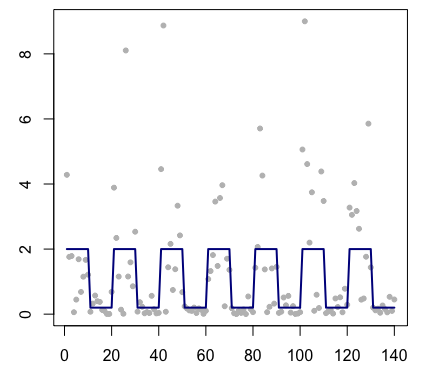}
         \caption{teeth-type} 
     \end{subfigure}
     \begin{subfigure}[b]{0.24\textwidth}
         \includegraphics[width=\textwidth]{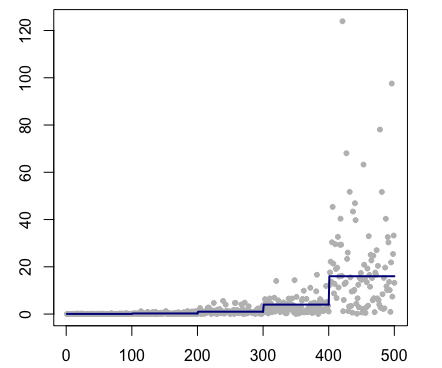}
         \caption{stairs-type} 
     \end{subfigure}
        \caption{(A) and (B): the test signals of the form $\exp(\bsg^{*})$ (solid line) and simulated data (dots) for Poisson model. (C) and (D): the test signals of the form of $1/\bsg^{*}$ (solid line) and simulated data (dots) for exponential model.} 
        \label{poi-exp-signal}
\end{figure}

To investigate the performance of ES and the candidates in $\widehat\gGamma$, we mimic the test signals \texttt{fms} and \texttt{mix} for the Poisson model and \texttt{teeth10} and \texttt{stairs10} for the exponential model. Additionally, we examine scenarios where outliers are present in the observations for the simulated signals \texttt{fms} (Poisson) and \texttt{teeth10} (exponential). The specific settings of these signals, along with the method for introducing outliers, are detailed in the Appendix. Figure~\ref{poi-exp-signal} displays the four underlying signals along with an example of the simulated data for each.

\begin{center}
{\small\renewcommand*{\arraystretch}{0.43}
\setlength\tabcolsep{1.8pt}
\begin{longtable}{ccc|cc>{\columncolor[gray]{0.8}}ccc|>{\columncolor[gray]{0.8}}c|c}
\caption{Frequencies of $\widehat N-N$ and $\widehat R_{n}(\cdot)$ of ES and its competitors for Poisson model over 1000 simulated sample paths.}\vspace{10pt}
\label{result-poi}\\
\hline\hline
 & & &\multicolumn{5}{c|}{$\widehat N-N$}&&\\
 \cline{4-8}
Method& Signal&Outlier& $\leq-2$& -1& 0& 1& $\geq2$& $\widehat R_{n}(\cdot)$&Contribution\\
\hline
ES&fms-type&No& 0.002& 0.050& \bf{0.878}& 0.062& 0.008&2.51\ $\pm$\ 0.09&-\\
SMUCE&fms-type&No&0.288& 0.528& 0.184& 0.000& 0.000&6.21\ $\pm$\ 0.19&0.184\\
CL&fms-type&No&0.000& 0.046& \bf{0.854}& 0.082&0.018&2.54\ $\pm$\ 0.10&0.725\\
$\mbox{CBS}_{t}+\rho$&fms-type&No&0.030& 0.254& 0.546& 0.131& 0.039&5.34\ $\pm$\ 0.11&0.000\\
$\mbox{cumSeg}_{t}+\rho$&fms-type&No& 0.424& 0.374& 0.193& 0.009& 0.000&7.26\ $\pm$\ 0.20&0.001\\
$\mbox{PELT}_{t}+\rho$&fms-type&No&0.003& 0.054&\bf{0.867}& 0.062& 0.014&2.56\ $\pm$\ 0.10&0.015\\
$\mbox{WBS sSIC}_{t}+\rho$&fms-type&No&0.010& 0.132& 0.781& 0.051&0.026&3.08\ $\pm$\ 0.12&0.013\\
$\mbox{FDR}_{t}(\alpha=0.05)+\rho$&fms-type&No&0.288& 0.528& 0.184& 0.000& 0.000&5.97\ $\pm$\ 0.18&0.000\\
$\mbox{robseg(Huber)}_{t}+\rho$&fms-type&No&0.001& 0.035&\bf{0.800}& 0.130& 0.034&2.68\ $\pm$\ 0.10&0.032\\
$\mbox{robseg(biweight)}_{t}+\rho$&fms-type&No& 0.005& 0.073&\bf{0.867}& 0.048& 0.007&2.63\ $\pm$\ 0.10&0.030\\
\hline
ES&fms-type&Yes& 0.001& 0.092& \bf{0.825}& 0.070& 0.012&3.78\ $\pm$\ 0.11&-\\
SMUCE&fms-type&Yes&0.000& 0.000& 0.000& 0.000& 1.000&12.76\ $\pm$\ 0.21&0.000\\
CL&fms-type&Yes&0.000& 0.000&0.000& 0.000&1.000&8.58\ $\pm$\ 0.11&0.000\\
$\mbox{CBS}_{t}+\rho$&fms-type&Yes&0.521& 0.354& 0.086& 0.035& 0.004&11.98\ $\pm$\ 0.36&0.000\\
$\mbox{cumSeg}_{t}+\rho$&fms-type&Yes& 0.795& 0.164& 0.038& 0.003& 0.000&12.01\ $\pm$\ 0.28&0.009\\
$\mbox{PELT}_{t}+\rho$&fms-type&Yes&0.000& 0.000& 0.000& 0.000& 1.000&8.45\ $\pm$\ 0.10&0.000\\
$\mbox{WBS sSIC}_{t}+\rho$&fms-type&Yes&0.000& 0.000& 0.000& 0.000& 1.000&8.82\ $\pm$\ 0.12&0.000\\
$\mbox{FDR}_{t}(\alpha=0.05)+\rho$&fms-type&Yes&0.000& 0.000& 0.000& 0.000& 1.000&11.35\ $\pm$\ 0.18&0.001\\
$\mbox{robseg(Huber)}_{t}+\rho$&fms-type&Yes&0.000& 0.008& 0.048& 0.062& 0.882&6.13\ $\pm$\ 0.13&0.053\\
$\mbox{robseg(biweight)}_{t}+\rho$&fms-type&Yes& 0.000& 0.092&\bf{0.839}& 0.066& 0.003&3.74\ $\pm$\ 0.11&0.937\\
\hline
ES&mix-type&No& 0.005& 0.371& \bf{0.523}& 0.091& 0.010&3.98\ $\pm$\ 0.09&-\\
SMUCE&mix-type&No&0.128& 0.828& 0.044& 0.000& 0.000&4.67\ $\pm$\ 0.13&0.339\\
CL&mix-type&No&0.014& 0.439&0.466& 0.071&0.010&3.99\ $\pm$\ 0.09&0.481\\
$\mbox{CBS}_{t}+\rho$&mix-type&No&0.034& 0.448& 0.358& 0.122& 0.038&13.39\ $\pm$\ 0.11&0.000\\
$\mbox{cumSeg}_{t}+\rho$&mix-type&No& 0.990& 0.010& 0.000& 0.000& 0.000&31.18\ $\pm$\ 0.45&0.000\\
$\mbox{PELT}_{t}+\rho$&mix-type&No&0.010& 0.443&0.466& 0.071& 0.010&4.03\ $\pm$\ 0.09&0.027\\
$\mbox{WBS sSIC}_{t}+\rho$&mix-type&No&0.018& 0.509& 0.402& 0.056&0.015&4.03\ $\pm$\ 0.09&0.013\\
$\mbox{FDR}_{t}(\alpha=0.05)+\rho$&mix-type&No&0.128& 0.828& 0.044& 0.000& 0.000&4.67\ $\pm$\ 0.12&0.000\\
$\mbox{robseg(Huber)}_{t}+\rho$&mix-type&No&0.003& 0.293&\bf{0.530}& 0.149& 0.025&4.15\ $\pm$\ 0.09&0.099\\
$\mbox{robseg(biweight)}_{t}+\rho$&mix-type&No& 0.014& 0.486& 0.458& 0.040& 0.002&4.06\ $\pm$\ 0.09&0.041\\
\hline\hline
\endfirsthead
\end{longtable}}
\end{center}

The results of Poisson model are shown in Table~\ref{result-poi}. Let us first comment the two existing estimators in the literature, namely SMUCE and CL. In both of the scenarios with or without outliers, the performance of SMUCE is quite poor at least under these two test signals. When no outlier presents in the observations, SMUCE has a tendency to underestimate the number of changepoints for both of the two signals \texttt{fms-type} and \texttt{mix-type}. When there are outliers, SMUCE is sensitive to them therefore overestimates the number of changepoints. The CL performs much better than SMUCE in the scenario where no outlier presents in the observations but it is also not robust with respect to the outliers. When no outlier presents, our estimator ES slightly improves the performance of CL on detecting changes under both of the two signals. When there are outliers presenting in the observations, ES obviously outperforms CL as a consequence of enjoying the excellent performance given by $\mbox{robseg (biweight)}_{t}$. Interestingly, we find that when there is no outlier presenting in the observations, the combinations $\mbox{PELT}_{t}+\rho$ and $\mbox{robseg}_{t}+\rho$ are perhaps nice choices at least under these two signals.

The results for exponential model are shown in Table~\ref{result-exp}. For the \texttt{teeth-type} signal without an outlier, the ES obviously outperforms any single candidate by selecting mainly from CL and $\mbox{robseg (biweight)}_{t}+\rho$. When there are outliers, $\mbox{robseg (biweight)}_{t}+\rho$ is the best one among all and we observe that ES improves the frequency of selecting $\mbox{robseg (biweight)}_{t}+\rho$ as the final estimator so that finally ES achieves a competitive performance compared to $\mbox{robseg (biweight)}_{t}+\rho$ and significantly outperforms the existing estimator CL. For the \texttt{stairs-type} signal, CL performs well, but ES shows a slight improvement by leveraging contributions from other candidates in $\widehat\gGamma$.

\begin{center}
{\small\renewcommand*{\arraystretch}{0.43}
\setlength\tabcolsep{1.4pt}
\begin{longtable}{ccc|cc>{\columncolor[gray]{0.8}}ccc|>{\columncolor[gray]{0.8}}c|c}
\caption{Frequencies of $\widehat N-N$ and $\widehat R_{n}(\cdot)$ of ES and its competitors for exponential model over 1000 simulated sample paths.}\vspace{10pt}
\label{result-exp}\\
\hline\hline
 & & &\multicolumn{5}{c|}{$\widehat N-N$}&&\\
 \cline{4-8}
Method& Signal&Outlier& $\leq-2$& -1& 0& 1& $\geq2$& $\widehat R_{n}(\cdot)$&Contribution\\
\hline
ES&teeth-type&No& 0.327& 0.077& \bf{0.468}& 0.106& 0.022&7.69\ $\pm$\ 0.25&-\\
CL&teeth-type&No&0.381& 0.055&0.411& 0.116&0.037&9.27\ $\pm$\ 0.38&0.766\\
$\mbox{SMUCE}_{t}+\rho$&teeth-type&No&0.998& 0.002& 0.000& 0.000& 0.000&20.29\ $\pm$\ 0.14&0.000\\
$\mbox{CBS}_{t}+\rho$&teeth-type&No&1.000& 0.000& 0.000& 0.000& 0.000&22.52\ $\pm$\ 0.06&0.000\\
$\mbox{cumSeg}_{t}+\rho$&teeth-type&No&1.000& 0.000& 0.000& 0.000& 0.000&22.56\ $\pm$\ 0.05&0.000\\
$\mbox{PELT}_{t}+\rho$&teeth-type&No&0.134& 0.068& 0.241& 0.191& 0.366&9.14\ $\pm$\ 0.19&0.015\\
$\mbox{WBS sSIC}_{t}+\rho$&teeth-type&No&0.829& 0.022& 0.058& 0.036&0.055&18.62\ $\pm$\ 0.37&0.007\\
$\mbox{FDR}_{t}(\alpha=0.05)+\rho$&teeth-type&No&0.998& 0.002& 0.000& 0.000& 0.000&20.30\ $\pm$\ 0.14&0.000\\
$\mbox{robseg(Huber)}_{t}+\rho$&teeth-type&No&0.076& 0.096& 0.263& 0.227& 0.338&8.42\ $\pm$\ 0.18&0.023\\
$\mbox{robseg(biweight)}_{t}+\rho$&teeth-type&No& 0.435& 0.122&0.348& 0.082& 0.013&8.86\ $\pm$\ 0.21&0.189\\
\hline
ES&teeth-type&Yes& 0.383& 0.082& \bf{0.303}& 0.151& 0.081&9.38\ $\pm$\ 0.25&-\\
CL&teeth-type&Yes&0.500& 0.048&0.169& 0.128&0.155&12.42\ $\pm$\ 0.42&0.534\\
$\mbox{SMUCE}_{t}+\rho$&teeth-type&Yes&1.000& 0.000& 0.000& 0.000& 0.000&22.02\ $\pm$\ 0.14&0.000\\
$\mbox{CBS}_{t}+\rho$&teeth-type&Yes&1.000& 0.000& 0.000& 0.000& 0.000&24.43\ $\pm$\ 0.07&0.001\\
$\mbox{cumSeg}_{t}+\rho$&teeth-type&Yes&1.000& 0.000& 0.000& 0.000& 0.000&24.49\ $\pm$\ 0.06&0.000\\
$\mbox{PELT}_{t}+\rho$&teeth-type&Yes&0.090& 0.069& 0.131& 0.162& 0.548&10.48\ $\pm$\ 0.18&0.023\\
$\mbox{WBS sSIC}_{t}+\rho$&teeth-type&Yes&0.908& 0.014& 0.017& 0.024&0.037&21.82\ $\pm$\ 0.32&0.008\\
$\mbox{FDR}_{t}(\alpha=0.05)+\rho$&teeth-type&Yes&1.000& 0.000& 0.000& 0.000& 0.000&22.02\ $\pm$\ 0.14&0.001\\
$\mbox{robseg(Huber)}_{t}+\rho$&teeth-type&Yes&0.090& 0.074& 0.202& 0.222& 0.412&9.37\ $\pm$\ 0.18&0.105\\
$\mbox{robseg(biweight)}_{t}+\rho$&teeth-type&Yes& 0.456& 0.106&\bf{0.316}& 0.105& 0.017&9.48\ $\pm$\ 0.20&0.328\\
\hline
ES&stairs-type&No& 0.000& 0.000& \bf{0.923}& 0.067& 0.010&2.09\ $\pm$\ 0.08&-\\
CL&stairs-type&No&0.000& 0.000& \bf{0.907}& 0.075& 0.018&2.10\ $\pm$\ 0.08&0.977\\
$\mbox{SMUCE}_{t}+\rho$&stairs-type&No&0.000& 0.008& 0.489& 0.225& 0.278&4.28\ $\pm$\ 0.19&0.003\\
$\mbox{CBS}_{t}+\rho$&stairs-type&No&0.006& 0.134& 0.594& 0.193& 0.073&6.27\ $\pm$\ 0.31&0.000\\
$\mbox{cumSeg}_{t}+\rho$&stairs-type&No&0.002& 0.120& 0.682& 0.192& 0.004&7.54\ $\pm$\ 0.32&0.000\\
$\mbox{PELT}_{t}+\rho$&stairs-type&No&0.000& 0.000& 0.032& 0.041& 0.927&6.56\ $\pm$\ 0.18&0.000\\
$\mbox{WBS sSIC}_{t}+\rho$&stairs-type&No&0.000& 0.003& 0.456& 0.094&0.447&4.63\ $\pm$\ 0.17&0.001\\
$\mbox{FDR}_{t}(\alpha=0.05)+\rho$&stairs-type&No&0.000& 0.008& 0.489& 0.225& 0.278&4.27\ $\pm$\ 0.19&0.002\\
$\mbox{robseg(Huber)}_{t}+\rho$&stairs-type&No&0.000& 0.000& 0.207& 0.144& 0.649&4.84\ $\pm$\ 0.16&0.001\\
$\mbox{robseg(biweight)}_{t}+\rho$&stairs-type&No& 0.000& 0.000&0.699& 0.183& 0.118&3.21\ $\pm$\ 0.12&0.016\\
\hline\hline
\endfirsthead
\end{longtable}}
\end{center}\vspace{-35pt}

\section{Real data applications}\label{realdata}
In this section, we apply our estimator selection procedure to two real datasets and evaluate its performance. The first dataset consists of DNA copy number observations from biological research, where the Gaussian model is used to detect changes. The second dataset is the British coal disasters dataset, for which the Poisson model is applied.

\subsection{Detecting changes in DNA copy numbers}
It is well known that in normal human cells, the number of DNA copies is two. As demonstrated in numerous biological studies (e.g., \cite{10.1093/hmg/ddg261} and \cite{articlenature}), the pathogenesis of several diseases, including various cancers and mental retardation, is often linked to chromosomal aberrations such as deletions, duplications, and/or amplifications. These chromosomal changes result in DNA copy numbers in affected regions differing from the normal count of two. Using techniques such as microarray and sequencing experiments, biologists have developed methods to measure DNA copy numbers for selected genes in genomes, recording the results as a sequence of observations $\bsY=(Y_{1},\ldots,Y_{n})$. The statistical challenge is to detect abrupt changes in the means of these observations. To address this, we apply the Gaussian model with an estimated variance.

The R package \texttt{jointseg} (\cite{jointseg}) provides two real datasets, GSE11976 and GSE29172, from which to resample, with the true changepoint locations already known. However, since we do not have the information about the true value of $\bsg^{*}$ on each segment, it is not possible to compute the pseudo Hellinger distance between each estimator and the true values. Note that for both GSE11976 and GSE29172 datasets, we need to choose the tumour fraction when resampling from them. We consider the tumour fraction levels 0.79 and 1 for the dataset GSE11976 and the levels 0.7 and 1 for GSE29172 which turns out to be the situations where the size of each jump at the changepoint is relatively large as indicated in Figure~9 of \cite{ChaDetectOut}. Therefore, we can roughly evaluate the performance of each estimator by its frequency of correctly estimating the number of changepoints. Although our selection procedure can be applied in the scenario where small amount of outliers present in the observations, as we have seen in Section~5.2 in the paper, some candidates in $\widehat\gGamma$ are sensitive to the outliers. To avoid the phenomenon that an estimator systematically underestimates the number of changepoints but due to the sensitivity to outliers it accidentally gives a correct number of segments, we run a smooth procedure on the data before applying all the estimation procedures by implementing the function \texttt{smooth.CNA} from the famous R package \texttt{DNAcopy}. Moreover, since we have seen in the simulation study that the performance of CBS and cumSeg is quite poor, we remove these two estimators from our candidates set $\widehat\gGamma$ for simplicity. For each dataset and each level of tumour fraction, we simulate 1000 profiles of length $n=1000$ with 5 changepoints where the length of each segment is at least 20. The results are shown in Table~\ref{result-dna}. As one can observe, among the state-of-art ones, robseg (biweight) is the best for correctly estimating the number of changepoints on this dataset. By running a data-driven procedure to select among the candidates set $\widehat\gGamma$, our selected estimator ES shows a competitive performance in this situation as compared to the best one robseg (biweight).

\begin{center}
{\small\renewcommand*{\arraystretch}{0.75}
\setlength\tabcolsep{2.2pt}
\begin{longtable}{ccc|ccc>{\columncolor[gray]{0.8}}cccc|c}
\caption{Frequencies of $\widehat N-N$ of ES and its competitors for DNA copy numbers data. Contribution denotes the frequency of each competitor being selected as ES. Bold: highest empirical frequency of $\widehat N-N=0$ and those with frequencies within 10\% off the highest.}\vspace{30pt}
\label{result-dna}\\
\hline\hline
& & &\multicolumn{7}{c|}{$\widehat N-N$}&\\
 \cline{4-10}
Method&Dataset&Fraction& $\leq-3$&-2& -1& 0& 1&2& $\geq3$&Contribution\\
\hline
ES&GSE11976 & 0.79& 0.003& 0.028&0.044&\bf{0.771}&0.108&0.031&0.015&-\\
PELT& GSE11976& 0.79&0.000&0.002&0.008&0.198&0.096&0.196&0.500&0.060\\
SMUCE&GSE11976 &0.79 & 0.004&0.021&0.124& 0.391& 0.203& 0.139&0.118&0.147\\
CL&GSE11976 &0.79 & 0.011& 0.066&0.053& 0.550& 0.117&0.118&0.085&0.393\\
WBS sSIC&GSE11976 &0.79 &0.005& 0.031&0.066& 0.508&0.066&0.174&0.150&0.100\\
FDR($\alpha=0.05$)&GSE11976 &0.79 & 0.000& 0.005&0.011& 0.096& 0.056&0.126&0.706&0.020\\
robseg(Huber)& GSE11976& 0.79& 0.001& 0.012& 0.022& 0.569& 0.193&0.110&0.093&0.121\\
robseg(biweight)& GSE11976& 0.79& 0.002& 0.046&0.045& \bf{0.778}& 0.102&0.019&0.008&0.159\\
\hline
ES&GSE11976 & 1.00& 0.000& 0.003&0.007&\bf{0.790}& 0.100& 0.046&0.054&-\\
PELT& GSE11976& 1.00&0.000&0.000&0.000&0.243&0.067&0.195&0.495&0.046\\
SMUCE&GSE11976 &1.00 & 0.000&0.002&0.035& 0.395& 0.178&0.177&0.213&0.208\\
CL&GSE11976 &1.00& 0.001& 0.011&0.011&0.604&0.098&0.170&0.105&0.357\\
WBS sSIC&GSE11976 &1.00 &0.000& 0.003&0.008&0.536&0.060&0.225&0.168&0.075\\
FDR($\alpha=0.05$)&GSE11976 &1.00 & 0.000& 0.000&0.004& 0.138& 0.059&0.126&0.673&0.018\\
robseg(Huber)& GSE11976&1.00& 0.000& 0.002& 0.004& 0.559& 0.163&0.126&0.146&0.155\\
robseg(biweight)& GSE11976&1.00& 0.000& 0.010&0.006&\bf{0.794}& 0.101&0.043&0.046&0.141\\
\hline
ES&GSE29172 & 0.70& 0.014& 0.136&0.133&\bf{0.596}& 0.088& 0.028&0.005&-\\
PELT& GSE29172& 0.70&0.003&0.027&0.054&0.210&0.139&0.181&0.386&0.089\\
SMUCE&GSE29172 &0.70 & 0.016& 0.112& 0.307& 0.247& 0.176& 0.087&0.055&0.099\\
CL&GSE29172 &0.70 & 0.035& 0.159&0.155& 0.305& 0.129&0.126&0.091&0.302\\
WBS sSIC&GSE29172 &0.70 &0.022& 0.105&0.155& 0.290&0.113&0.173&0.142&0.046\\
FDR($\alpha=0.05$)&GSE29172 &0.70 & 0.003& 0.024&0.075& 0.133& 0.112&0.133&0.520&0.032\\
robseg(Huber)& GSE29172& 0.70& 0.007& 0.068& 0.087& 0.533& 0.163&0.092&0.050&0.224\\
robseg(biweight)& GSE29172& 0.70& 0.018& 0.168&0.153&\bf{0.597}& 0.052&0.012&0.000&0.208\\
\hline
ES&GSE29172 & 1.00& 0.000& 0.005&0.003& \bf{0.828}&0.093&0.051& 0.020&-\\
PELT& GSE29172& 1.00&0.000&0.001&0.001&0.233&0.070&0.251&0.444&0.046\\
SMUCE&GSE29172 &1.00 & 0.000& 0.004& 0.044&0.416& 0.193& 0.199& 0.144&0.185\\
CL&GSE29172 &1.00 & 0.001& 0.009&0.006& 0.684& 0.077&0.163&0.060&0.427\\
WBS sSIC&GSE29172 &1.00 &0.000& 0.006&0.009& 0.576&0.051&0.230&0.128&0.070\\
FDR($\alpha=0.05$)&GSE29172 &1.00 & 0.000& 0.001& 0.002& 0.119& 0.063&0.133&0.682&0.018\\
robseg(Huber)& GSE29172&1.00& 0.000& 0.001& 0.001& 0.594& 0.145&0.158&0.101&0.120\\
robseg(biweight)& GSE29172& 1.00& 0.000&0.007& 0.006&\bf{0.833}& 0.098& 0.043&0.013&0.134\\
\hline\hline
\endfirsthead
\end{longtable}}
\end{center}

\subsection{British coal disasters dataset}
The British coal disasters dataset is well-known in the context of Poisson segmentation (see, for example, \cite{10.1093/biomet/82.4.711}, \cite{doi:10.1198/106186001317243449}, \cite{articleFear06}, and \cite{10.5555/3045118.3045311}). We choose this dataset for two main reasons: First, the changepoints have been studied using various methods, making it easier to compare our results. Second, the sequence generally shows a decreasing trend over time, which may correlate with the implementation of safety regulations. Although rough, this provides some evidence for evaluating changepoint detection procedures on the dataset.

\begin{figure}[h]
        \begin{subfigure}{\linewidth}
        \includegraphics[width=\textwidth,height=5cm]{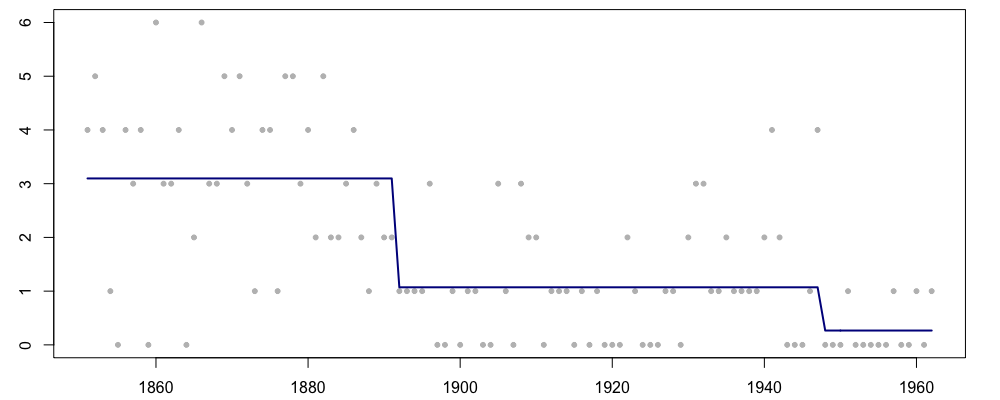}
     \end{subfigure}
       \caption{Coal mining disasters data (dots) and ES estimator (solid line).}
    \label{ukcoal}
\end{figure}

The data at hand include the number of each year coal disasters in UK during the period from March 15th, 1851 to March 22nd, 1962 with length $n=112$. In this situation, to detect changes along the sequence, Poisson model is considered together with the candidates set \eref{poi-gam} described in Section~\ref{poi-exp-sim}. We conclude the results of different estimators as follows. Concerning to the changepoints, there are in total three suggestions:
\begin{enumerate}[label=(\roman*)]
\item 1 changepoint at the year 1891: $\mbox{cumSeg}_{t}+\rho$, $\mbox{PELT}_{t}+\rho$, $\mbox{WBS sSIC}_{t}+\rho$, $\mbox{FDR}_{t}(\alpha=0.05)+\rho$ and $\mbox{robseg(biweight)}_{t}+\rho$;
\item 2 changepoints at the year 1891 and 1947: $\SMUCE$ and $\CCL$;
\item 3 changepoints at the year 1891, 1929 and 1942: $\mbox{robseg(Huber)}_{t}+\rho$.
\end{enumerate}
Our selection procedure finally choose SMUCE as ES, i.e. we support the suggestion with two changepoints at the year 1891 and 1947. The dataset as well as the result of ES (SMUCE) is plotted in Figure~\ref{ukcoal}.

Now we comment our result by comparing it with the existing ones in the literature. In \cite{10.1093/biomet/82.4.711}, they used the coal mining disasters data recorded per day and proposed a reversible jump MCMC approach to detect changepoints and estimate the intensity function. According to the Figure~2 in the same paper, the model with two changepoints has the highest posterior probability. Moreover, according to their Figure~3, in the two changepoints scenario, the posterior mode is approximately 14,000 days for the first changepoint and 35,000 days for the second one. This is very close to our result since counting from March 15th, 1851, 14,000 days is between the year 1889 and 1890 and 35,000 days is the time between the year 1946 and 1947. Later, a Bayesian binary segmentation procedure was proposed by \cite{doi:10.1198/106186001317243449} to locate changepoints for Poisson process. Based on two different tests they adopted, their procedure obtained two different sets of changepoints (one changepoint for applying Bayes factor criterion and two for applying BIC approximation criterion) where the locations of changepoints for these two models are quite similar to the results (i) and (ii) mentioned in the last paragraph. On the other hand, as it was pointed out in \cite{10.5555/3045118.3045311}, UK parliament passed several acts to improve the safety of mine works including the Coal Mines Regulation Acts of 1872 and 1887 and a further one in 1954 with mines and quarries acts. In general, it is reasonable to have a non-increasing expectation of the number of disasters after the year releasing these regulations. As it is shown in Figure~\ref{ukcoal}, the model with two changepoints meets the releasing regulation years 1887 and 1954. Considering the best fit with the time of released regulations and the results given in the literatures, we believe the two changepoints model for this dataset is the most reasonable one to the truth. 

\section*{Acknowledgements}
The author gratefully acknowledges the EU's Horizon 2020 programme No. 811017 and the NWO Vidi grant VI.Vidi.192.021.

\clearpage
\appendix
\setcounter{equation}{0}
\renewcommand{\theequation}{\arabic{equation}{$^*$}}
\section{Proofs}\label{proofs}
We first introduce some notation for later use. Recall that $(\sX,\cX)=(\sW\times\sY,\cW\otimes\cY)$. We denote $\sbP$ the set of all product probabilities on $(\sX^{n},\cX^{\otimes n})$. For all $i\in\{1,\ldots,n\}$, we denote the true distribution of $X_{i}=(W_{i},Y_{i})$ by $P_{i}^{*}$ and denote the true joint distribution of $\bsX=(X_{1},\ldots,X_{n})$ by $\gP^{*}=\otimes_{i=1}^{n}P_{i}^{*}\in\sbP$. We denote $\gP_{\bsg}=\otimes_{i=1}^{n}P_{i,\bsg}$ the joint distribution of independent random variables $(W_{1},Y_{1}),\ldots,(W_{n},Y_{n})$ for which the conditional distribution of $Y_{i}$ given $W_{i}=w_{i}$ is given by $R_{\bsg(w_{i})}\in\sQ_{0}$ for all $i\in\{1,\ldots,n\}$. Under such a notation setting, we have $P_{i}^{*}=R_{i}^{*}\cdot P_{W_{i}}$, $P_{i,\bsg}=R_{\bsg}\cdot P_{W_{i}}$ as well as the following equality
\begin{equation}\label{l1}
h^{2}(P_{i}^{*},P_{i,\bsg})=\int_{\sW}h^{2}(R_{i}^{*}(w),R_{\bsg(w)})dP_{W_{i}}(w).
\end{equation}
We define the pseudo Hellinger distance $\gh$ between two probabilities $\gP=\otimes_{i=1}^{n}P_{i}$ and $\gP'=\otimes_{i=1}^{n}P'_{i}$ by 
\begin{equation}\label{l2}
\gh^{2}(\gP,\gP')=\sum_{i=1}^{n}h^{2}(P_{i},P'_{i}).
\end{equation}
As an immediate consequence of \eref{l1} and \eref{l2}, for any $\bsg\in\widetilde\gGamma$,

\begin{align}
\gh^{2}(\gR^{*},\gR_{\bsg})&=\sum_{i=1}^{n}\int_{\sW}h^{2}(R_{i}^{*}(w),R_{\bsg(w)})dP_{W_{i}}(w)\nonumber\\
&=\sum_{i=1}^{n}h^{2}(P_{i}^{*},P_{i,\bsg})=\gh^{2}(\gP^{*},\gP_{\bsg}).\label{l3}
\end{align}
For each $m\in\cM$, we define the set of probabilities $\sbP_{m}=\{\gP_{\bsg},\;\bsg\in\gGamma_{m}\}$ and $\widetilde\sbP=\{\gP_{\bsg},\;\bsg\in\widetilde\gGamma\}$ with $\widetilde\gGamma=\cup_{m\in\cM}\gGamma_{m}$. For any $y>0$, $\gP^{*}\in\sbP$, $\sbP_{m_{1}}, \sbP_{m_{2}}$ with $m_{1}, m_{2}\in\cM$, we define the set 
\begin{align*}
&\sbB^{\sbP_{m_{1}}\times\sbP_{m_{2}}}(\gP^{*}, y)\\
=&\left\{(\gP_{\bsg_{1}}, \gP_{\bsg_{2}})\big|\ \gP_{\bsg_{1}}\in\sbP_{m_{1}}, \gP_{\bsg_{2}}\in\sbP_{m_{2}}, \gh^{2}(\gP^{*}, \gP_{\bsg_{1}})+\gh^{2}(\gP^{*}, \gP_{\bsg_{2}})<y^{2}\right\}
\end{align*}
and for any $\bsg_{1},\bsg_{2}\in\widetilde\gGamma$, we set $$\gZ(\gX, \bsg_{1}, \bsg_{2})=\gT(\gX, \bsg_{1}, \bsg_{2})-\E\cro{\gT(\gX, \bsg_{1}, \bsg_{2})}.$$ We then introduce bellow Proposition 45 of \cite{MR3595933} which is an extensional version of Talagrand's Theorem on the supremum of empirical processes proved in \cite{MR2319879}.
\begin{prop}\label{extension-talagrand}
Let $T$ be some finite or countable set, $U_{1}$,...,$U_{n}$ be independent centered random vectors with values in $\R^{T}$ and let $$Z=\sup_{t\in T}\Big|\sum_{i=1}^{n}U_{i,t}\Big|.$$ If for some positive numbers $b$ and $v$,
$$\max_{i=1,...,n}|U_{i,t}|\leq b \text{\quad and\quad} \sum_{i=1}^{n}\E\cro{U_{i,t}^{2}}\leq v^{2} \text{\quad for all\ } t\in T,$$ then, for all positive numbers $c$ and $x$,
\[
\P\cro{Z\leq(1+c)\E(Z)+(8b)^{-1}cv^{2}+2(1+8c^{-1})bx}\geq1-e^{-x}.
\]
\end{prop}
\subsection{Elementary results and proofs}
Before proving our main theorem, we first present two preliminary results and their proofs in this section.
\begin{lem}\label{lem-cle}
Let $m_{1},m_{2}\in\cM$ be any two partitions on $\sW$. The class of functions $$\sF(m_{1},m_{2})=\left\{\frac{r_{\bsg_{2}}}{r_{\bsg_{1}}}:(w,y)\mapsto \frac{r_{\bsg_{2}(w)}(y)}{r_{\bsg_{1}(w)}(y)},\; \bsg_{1}\in\gGamma_{m_{1}},\; \bsg_{2}\in\gGamma_{m_{2}}\right\}$$ on $\sX=\sW\times\sY$ is a VC-subgraph class with dimension not larger than $2|m_{1}\vee m_{2}|+1$.
\end{lem}
\begin{proof}
For any $\bsg_{1}\in\gGamma_{m_{1}}$ and $\bsg_{2}\in\gGamma_{m_{2}}$, we define function $g_{\bsg_{1},\bsg_{2}}$ on $\sW\times\sY$ as
\begin{equation*}
g_{\bsg_{1},\bsg_{2}}(w,y)=T(y)\cro{u(\bsg_{2}(w))-u(\bsg_{1}(w))}-\cro{A(\bsg_{2}(w))-A(\bsg_{1}(w))}
\end{equation*}
and define $\sG(m_{1},m_{2})$ the class of functions as $$\sG(m_{1},m_{2})=\left\{g_{\bsg_{1}, \bsg_{2}}|\ \bsg_{1}\in\gGamma_{m_{1}}, \bsg_{2}\in\gGamma_{m_{2}}\right\}.$$
With the fact that $\sF(m_{1},m_{2})=\left\{e^{g},\;g\in\sG(m_{1},m_{2})\right\}$ and the exponential function is monotone on $\R$, by \cite{MR3595933} [Proposition~42-(ii)], it is enough to prove the conclusion holds for the class $\sG(m_{1},m_{2})$.

Let $K=|m_{1}\vee m_{2}|$ be the number of segments given by the refined partition $m_{1}\vee m_{2}$ and $\cI_{1},\ldots,\cI_{K}$ the resulted segments on $\sW$. For any $\bsg_{1}\in\gGamma_{m_{1}}$, we can rewrite it as $$\bsg_{1}(w)=\sum_{k=1}^{K}a_{k}\mathbbm{1}_{\cI_{k}}(w),\mbox{\quad where\ }(a_{1},\ldots,a_{K})\in I^{K}$$ and any $\bsg_{2}\in\gGamma_{m_{2}}$, $$\bsg_{2}(w)=\sum_{k=1}^{K}b_{k}\mathbbm{1}_{\cI_{k}}(w),\mbox{\quad where\ }(b_{1},\ldots,b_{K})\in I^{K}.$$ As an immediate consequence, for any $g_{\bsg_{1},\bsg_{2}}\in\sG(m_{1},m_{2})$, it can be rewritten as $$g_{\bsg_{1},\bsg_{2}}(w,y)=\sum_{k=1}^{K}\cro{u(b_{k})-u(a_{k})}\mathbbm{1}_{\cI_{k}}(w)T(y)-\sum_{k=1}^{K}\cro{A(b_{k})-A(a_{k})}\mathbbm{1}_{\cI_{k}}(w).$$ Therefore, $\sG(m_{1},m_{2})$ is contained in a $2K$-dimensional vector space spanned by $$\left\{\mathbbm{1}_{\cI_{k}}(w), T(y)\mathbbm{1}_{\cI_{k}}(w),\;k=1,\ldots,K\right\}.$$ By Lemma 2.6.15 of \cite{MR1385671}, we conclude $\sG(m_{1},m_{2})$ is VC-subgraph on $\sX=\sW\times\sY$ with dimension not larger than $2K+1$.
\end{proof}
\begin{prop}\label{bound-by-y}
Let $m_{1},m_{2}\in\cM$ be any two partitions on $\sW$. Under Assumption~\ref{sum-dim}, for any $\gP^{*}\in\sbP$, $\eta\geq1$ and any $y>0$ satisfying
\begin{align*}
y^{2}&\geq\eta\cro{D_{n}(m_{1})+D_{n}(m_{2})},
\end{align*}
we have
\begin{align*}
\E\cro{\sup_{(\gP_{\bsg_{1}}, \gP_{\bsg_{2}})\in\sbB^{\sbP_{m_{1}}\times\sbP_{m_{2}}}(\gP^{*}, y)}\left|\gZ(\gX, \bsg_{1}, \bsg_{2})\right|}&\leq\cro{9.77\sqrt{\frac{2\alpha+1/2}{\eta}}+\frac{90(2\alpha+1/2)}{\eta}}y^{2}.
\end{align*}
\end{prop}
\begin{proof}
We set ${\bm{\tau}}=\otimes_{i=1}^{n}\tau_{i}$ with $\tau_{i}=P_{W_{i}}\otimes\nu$. For any $\bsg\in\widetilde\gGamma$, we denote $\gr_{\bsg}$ a density on $\sX^{n}=(\sW\times\sY)^{n}$ as
\begin{equation*}
\gr_{\bsg}(x_{1},\ldots,x_{n})=r_{\bsg}(x_{1})\cdots r_{\bsg}(x_{n}),\mbox{\quad for all\ }(x_{1},\ldots,x_{n})\in\sX^{n}
\end{equation*}
so that for any $\bsg\in\widetilde\gGamma$, we have $\gP_{\bsg}=\gr_{\bsg}\cdot{\bm{\tau}}$. For any $y>0$, we define $\sF_{y}(m_{1},m_{2})$ the class of functions on $\sX$ as
\[
\ac{\left.\psi\pa{\sqrt{\frac{r_{\bsg_{2}}}{r_{\bsg_{1}}}}}\right|\;\bsg_{1}\in\gGamma_{m_{1}}, \bsg_{2}\in\gGamma_{m_{2}}, \gh^{2}(\gP^{*},\gr_{\bsg_{1}}\cdot{\bm{\tau}})+\gh^{2}(\gP^{*},\gr_{\bsg_{2}}\cdot{\bm{\tau}})< y^{2}}.
\]
Since $\sF_{y}(m_{1},m_{2})$ is a subset of the collection $$\ac{\left.\psi\left(\sqrt{\frac{r_{\bsg_{2}}}{r_{\bsg_{1}}}}\right)\right|\; \bsg_{1}\in\gGamma_{m_{1}}, \bsg_{2}\in\gGamma_{m_{2}}}$$ and the function $\psi$ is monotone, it follows from Lemma~\ref{lem-cle} and Proposition~42-(ii) of \cite{MR3595933} that $\sF_{y}(m_{1},m_{2})$ is VC-subgraph on $\sX$ with dimension not larger than $\overline V=2|m_{1}\vee m_{2}|+1$. Besides, by Proposition~3 of \cite{BarBir2018}, our choice of the function $\psi$ satisfies their Assumption~2 and more precisely (11) in their paper with $a_{2}^{2}=3\sqrt{2}$ so that for any $y>0$, 
\begin{equation}\label{b-emp}
\sup_{f\in\sF_{y}(m_{1},m_{2})}n^{-1}\sum_{i=1}^{n}\E\cro{f^{2}(X_{i})}\leq\frac{a_{2}^{2}y^{2}}{n}.
\end{equation}
Moreover, since the function $\psi$ takes values in $\cro{-1,1}$, we derive from \eref{b-emp} that $$\sup_{f\in\sF_{y}(m_{1},m_{2})}n^{-1}\sum_{i=1}^{n}\E\cro{f^{2}(X_{i})}\leq\left(\frac{a_{2}^{2}y^{2}}{n}\right)\wedge 1\leq1.$$ To bound the expectation of the supremum of an empirical process over a VC-subgraph class, we apply Theorem~2 of \cite{Baraud2020} to $\sF_{y}(m_{1},m_{2})$ and obtain
\begin{align}
&\ \E\cro{\sup_{(\gP_{\bsg_{1}}, \gP_{\bsg_{2}})\in\sbB^{\sbP_{m_{1}}\times\sbP_{m_{2}}}(\gP^{*}, y)}\left|\gZ(\gX, \bsg_{1}, \bsg_{2})\right|}\nonumber\\
=&\ \E\cro{\sup_{(\gP_{\bsg_{1}}, \gP_{\bsg_{2}})\in\sbB^{\sbP_{m_{1}}\times\sbP_{m_{2}}}(\gP^{*}, y)}\left|\gT(\gX, \bsg_{1}, \bsg_{2})-\E\cro{\gT(\gX, \bsg_{1}, \bsg_{2})}\right|}\nonumber\\
=&\ \E\cro{\sup_{f\in\sF_{y}(m_{1},m_{2})}\ab{\sum_{i=1}^{n}\left(f(X_{i})-\E\cro{f(X_{i})}\right)}}\nonumber\\
\leq&\ 9.77y\sqrt{\overline VL_{n}(y)}+90\overline VL_{n}(y),\label{direct-y}
\end{align}
where $L_{n}(y)=9.11+\log_{+}\cro{n/\left(3\sqrt{2}y^{2}\right)}.$ Under Assumption~\ref{sum-dim}, there exists a constant $\alpha\geq1$ such that 
\begin{equation}\label{ass-vc}
\overline V=2|m_{1}\vee m_{2}|+1\leq 2\alpha(|m_{1}|+|m_{2}|)+1\leq\left(2\alpha+\frac{1}{2}\right)(|m_{1}|+|m_{2}|).
\end{equation}
Therefore, combining \eref{direct-y} and \eref{ass-vc}, we obtain
\begin{align}
&\E\cro{\sup_{(\gP_{\bsg_{1}}, \gP_{\bsg_{2}})\in\sbB^{\sbP_{m_{1}}\times\sbP_{m_{2}}}(\gP^{*}, y)}\left|\gZ(\gX, \bsg_{1}, \bsg_{2})\right|}\nonumber\\
\leq&9.77y\sqrt{\left(2\alpha+\frac{1}{2}\right)\left(|m_{1}|+|m_{2}|\right)L_{n}(y)}+90\left(2\alpha+\frac{1}{2}\right)(|m_{1}|+|m_{2}|)L_{n}(y).\label{em-risk}
\end{align}
Recall that $D_{n}(m)=|m|\cro{9.11+\log_{+}\left(n/|m|\right)}$. For any $\eta\geq1$, provided $$y^{2}\geq\eta\cro{D_{n}(m_{1})+D_{n}(m_{2})},$$ on the one hand, we have
\begin{align}
y^{2}&\geq\eta |m_{1}|\left(9.11+\log_{+}\pa{\frac{n}{|m_{1}|+|m_{2}|}}\right)\nonumber\\
&\quad+\eta|m_{2}|\left(9.11+\log_{+}\pa{\frac{n}{|m_{1}|+|m_{2}|}}\right)\nonumber\\
&=\eta(|m_{1}|+|m_{2}|)\cro{9.11+\log_{+}\pa{\frac{n}{|m_{1}|+|m_{2}|}}}\label{b-1}. 
\end{align}
On the other hand, (\ref{b-1}) also implies $y^{2}\geq |m_{1}|+|m_{2}|$. Therefore,
\begin{align}
L_{n}(y)&=9.11 +\log_{+}\pa{\frac{n}{3\sqrt{2}y^{2}}}\leq 9.11+\log_{+}\cro{\frac{n}{3\sqrt{2}(|m_{1}|+|m_{2}|)}}\nonumber \\
&\leq 9.11+\log_{+}\pa{\frac{n}{|m_{1}|+|m_{2}|}}.\label{L-bound}
\end{align}
Plugging (\ref{b-1}) and (\ref{L-bound}) into \eref{em-risk}, we complete the proof.
\end{proof}

\subsection{Proof of Theorem~\ref{thm-1}}
The proof of Theorem~\ref{thm-1} is inspired by the proof of Theorem~A.1 in \cite{BarBir2018}. Before we start to prove Theorem~\ref{thm-1}, we first show the following result.
\begin{prop}\label{empirical-process-bound}
Let numbers $a, \eta\geq1$ and $\delta, \vartheta>1$ such that
\begin{equation}\label{delta-upsilon-bound}
2\exp(-\vartheta)+\sum_{j=1}^{+\infty}\exp(-\vartheta\delta^{j})\leq1.
\end{equation}
Under Assumptions~\ref{weight-inequality} and \ref{sum-dim}, for any $\xi>0$ and for all $m_{1}, m_{2}\in\cM$ simultaneously, with probability at least $1-\Sigma^{2}e^{-\xi}$,
\begin{align*}
&\sup_{(\gP_{\bsg_{1}}, \gP_{\bsg_{2}})\in\sbP_{m_{1}}\times\sbP_{m_{2}}}\cro{\left|\gZ(\gX, \bsg_{1}, \bsg_{2})\right|-k_{1}\cro{\gh^{2}(\gP^{*}, \gP_{\bsg_{1}})+\gh^{2}(\gP^{*}, \gP_{\bsg_{2}})}}\\
&\quad\quad\quad\quad\quad\quad\leq k_{0}a\left\{\eta\cro{D_{n}(m_{1})+D_{n}(m_{2})}\vee\left(\Delta(m_{1})+\Delta(m_{2})+\vartheta+\xi\right)\right\},
\end{align*}
where 
\begin{align*}
k_{0}=&16\sqrt{\frac{9.77\sqrt{\frac{2\alpha+1/2}{\eta}}+\frac{90(2\alpha+1/2)}{\eta}+\frac{3\sqrt{2}}{16}}{2a}}+\frac{4}{a}\\
&+\left(9.77\sqrt{\frac{2\alpha+1/2}{\eta}}+\frac{90(2\alpha+1/2)}{\eta}\right),
\end{align*}
\begin{align*}
k_{1}=&16\sqrt{\frac{\delta\left(9.77\sqrt{\frac{2\alpha+1/2}{\eta}}+\frac{90(2\alpha+1/2)}{\eta}+\frac{3\sqrt{2}}{16}\right)}{2a}}+\frac{4}{a}\\
&+\left(9.77\sqrt{\frac{2\alpha+1/2}{\eta}}+\frac{90(2\alpha+1/2)}{\eta}\right)\delta.
\end{align*}
\end{prop}

\begin{proof}
Let $\xi>0$, $\delta, \vartheta>1$, $a, \eta\geq1$ and $m_{1},m_{2}\in\cM$ be fixed. For each $j\in\N$, we set $$x_{0}(m_{1}, m_{2})=\eta\left(D_{n}(m_{1})+D_{n}(m_{2})\right)\vee\left(\Delta(m_{1})+\Delta(m_{2})+\vartheta+\xi\right),$$ $$x_{j}(m_{1}, m_{2})=\delta^{j}x_{0}(m_{1}, m_{2}),\quad\quad y_{j}^{2}(m_{1}, m_{2})=ax_{j}(m_{1}, m_{2}).$$ For each $j\in\N$, we define the set 
\begin{align*}
&\cB_{j}^{\sbP_{m_{1}}\times\sbP_{m_{2}}}(\gP^{*})\\
=&\left\{(\gP_{\bsg_{1}},\gP_{\bsg_{2}})\in\sbP_{m_{1}}\times\sbP_{m_{2}}\big|\ y_{j}^{2}\leq\gh^{2}(\gP^{*}, \gP_{\bsg_{1}})+\gh^{2}(\gP^{*}, \gP_{\bsg_{2}})<y_{j+1}^{2}\right\}
\end{align*}
and set $$\cZ_{j}^{\sbP_{m_{1}}\times\sbP_{m_{2}}}(\gX)=\sup_{(\gP_{\bsg_{1}},\gP_{\bsg_{2}})\in\cB_{j}^{\sbP_{m_{1}}\times\sbP_{m_{2}}}(\gP^{*})}\left|\gZ(\gX, \bsg_{1}, \bsg_{2})\right|.$$ For simplifying the notations, let us drop the dependancy of $x_{j}$ and $y_{j}$ with respect to $m_{1}, m_{2}$ for a while. Since $\cB_{j}^{\sbP_{m_{1}}\times\sbP_{m_{2}}}(\gP^{*})\subset\sbB^{\sbP_{m_{1}}\times\sbP_{m_{2}}}(\gP^{*}, y_{j+1})$ and $y_{j+1}^{2}> y_{0}^{2}=ax_{0}\geq\eta\cro{D_{n}(m_{1})+D_{n}(m_{2})}$, under Assumption~\ref{sum-dim}, applying Proposition~\ref{bound-by-y} yields, 
\begin{align}
\E\cro{\cZ_{j}^{\sbP_{m_{1}}\times\sbP_{m_{2}}}(\gX)}&=\E\cro{\sup_{(\gP_{\bsg_{1}},\gP_{\bsg_{2}})\in\cB_{j}^{\sbP_{m_{1}}\times\sbP_{m_{2}}}(\gP^{*})}\left|\gZ(\gX, \bsg_{1}, \bsg_{2})\right|}\nonumber\\
&\leq\E\cro{\sup_{(\gP_{\bsg_{1}},\gP_{\bsg_{2}})\in\sbB^{\sbP_{m_{1}}\times\sbP_{m_{2}}}(\gP^{*}, y_{j+1})}\left|\gZ(\gX, \bsg_{1}, \bsg_{2})\right|}\nonumber\\
&\leq\left(9.77\sqrt{\frac{2\alpha+1/2}{\eta}}+\frac{90(2\alpha+1/2)}{\eta}\right)y_{j+1}^{2}.\label{sup-expectation-jth-ball}
\end{align}
For $i\in\{1,\ldots,n\}$, we set
\begin{equation}\label{Ui}
U_{i,(r_{\bsg_{1}},r_{\bsg_{2}})}=\psi\left(\sqrt{\frac{r_{\bsg_{2}(W_{i})}(Y_{i})}{r_{\bsg_{1}(W_{i})}(Y_{i})}}\right)-\E\cro{\psi\left(\sqrt{\frac{r_{\bsg_{2}(W_{i})}(Y_{i})}{r_{\bsg_{1}(W_{i})}(Y_{i})}}\right)}.
\end{equation}
With the fact that $\psi$ takes values in $\cro{-1,1}$, it is easy to observe that $$\max_{i=1,\ldots,n}\left|U_{i,(r_{\bsg_{1}},r_{\bsg_{2}})}\right|\leq2.$$ Moreover, $\psi$ satisfies the Assumption~2 more precisely (11) in \cite{BarBir2018} with $a_{2}^{2}=3\sqrt{2}$, we derive for each $j\in\N$, all $\bsg_{1}\in\gGamma_{m_{1}}$, $\bsg_{2}\in\gGamma_{m_{2}}$ such that $(\gP_{\bsg_{1}},\gP_{\bsg_{2}})\in\cB_{j}^{\sbP_{m_{1}}\times\sbP_{m_{2}}}(\gP^{*})$
\begin{align*}
\sum_{i=1}^{n}\E\cro{U^{2}_{i,(r_{\bsg_{1}},r_{\bsg_{2}})}}&\leq\sum_{i=1}^{n}\E\cro{\psi^{2}\left(\sqrt{\frac{r_{\bsg_{2}(W_{i})}(Y_{i})}{r_{\bsg_{1}(W_{i})}(Y_{i})}}\right)}\leq 3\sqrt{2}y_{j+1}^{2}.
\end{align*}
Then, for each $j\in\N$, we can apply Proposition~\ref{extension-talagrand} with $b=2$, $v^{2}=3\sqrt{2}y_{j+1}^{2}$ and $T=\cB_{j}^{\sbP_{m_{1}}\times\sbP_{m_{2}}}(\gP^{*})$ and obtain that for all $c>0$ and for all $(\gP_{\bsg_{1}},\gP_{\bsg_{2}})\in\cB_{j}^{\sbP_{m_{1}}\times\sbP_{m_{2}}}(\gP^{*})$ with probability at least $1-e^{-x_{j}}$,
\begin{align*}
\left|\gZ(\gX, \bsg_{1}, \bsg_{2})\right|\leq&\ \cZ_{j}^{\sbP_{m_{1}}\times\sbP_{m_{2}}}(\gX)\\
\leq&\ (1+c)\E\cro{\cZ_{j}^{\sbP_{m_{1}}\times\sbP_{m_{2}}}(\gX)}+\frac{3\sqrt{2}y_{j+1}^{2}c}{16}+4\left(1+\frac{8}{c}\right)x_{j}\\
\leq&\ (1+c)\left(9.77\sqrt{\frac{2\alpha+1/2}{\eta}}+\frac{90(2\alpha+1/2)}{\eta}\right)y_{j+1}^{2}\\&\ +\frac{3\sqrt{2}y_{j+1}^{2}c}{16}+4\left(1+\frac{8}{c}\right)x_{j}\\
\leq&\ (1+c)\left(9.77\sqrt{\frac{2\alpha+1/2}{\eta}}+\frac{90(2\alpha+1/2)}{\eta}\right)y_{j+1}^{2}\\&\ +\frac{3\sqrt{2}y_{j+1}^{2}c}{16}+\frac{4}{a}\left(1+\frac{8}{c}\right)y_{j}^{2}\\
\leq&\ (1+c)\left(9.77\sqrt{\frac{2\alpha+1/2}{\eta}}+\frac{90(2\alpha+1/2)}{\eta}\right)\delta y_{j}^{2}\\&+\cro{\frac{3\sqrt{2}c\delta}{16}+\frac{4}{a}\left(1+\frac{8}{c}\right)}y_{j}^{2}.
\end{align*}
Taking $$c=\sqrt{\frac{32}{\left(9.77\sqrt{\frac{2\alpha+1/2}{\eta}}+\frac{90(2\alpha+1/2)}{\eta}+\frac{3\sqrt{2}}{16}\right)\delta a}}$$ to minimize the bracketed term yields for all $(\gP_{\bsg_{1}},\gP_{\bsg_{2}})\in\cB_{j}^{\sbP_{m_{1}}\times\sbP_{m_{2}}}(\gP^{*})$, with probability at least $1-e^{-x_{j}}$
$$\left|\gZ(\gX, \bsg_{1}, \bsg_{2})\right|\leq k_{1}y_{j}^{2}.$$
By the definition of $\cB_{j}^{\sbP_{m_{1}}\times\sbP_{m_{2}}}(\gP^{*})$, we get for all $(\gP_{\bsg_{1}},\gP_{\bsg_{2}})$ belonging to $\cB_{j}^{\sbP_{m_{1}}\times\sbP_{m_{2}}}(\gP^{*})$, with probability at least $1-e^{-x_{j}}$,
$$\left|\gZ(\gX, \bsg_{1}, \bsg_{2})\right|\leq k_{1}y_{j}^{2}\leq k_{1}\cro{\gh^{2}(\gP^{*},\gP_{\bsg_{1}})+\gh^{2}(\gP^{*},\gP_{\bsg_{2}})}.$$
We define $$\cZ^{\sbP_{m_{1}}\times\sbP_{m_{2}}}(\gX)=\sup_{(\gP_{\bsg_{1}},\gP_{\bsg_{2}})\in\sbB^{\sbP_{m_{1}}\times\sbP_{m_{2}}}(\gP^{*}, y_{0})}\left|\gZ(\gX, \bsg_{1}, \bsg_{2})\right|.$$ With an analogous argument by applying Proposition~\ref{extension-talagrand} to $\cZ^{\sbP_{m_{1}}\times\sbP_{m_{2}}}(\gX)$ with $x=x_{0}$, we can obtain for all $(\gP_{\bsg_{1}},\gP_{\bsg_{2}})\in\sbB^{\sbP_{m_{1}}\times\sbP_{m_{2}}}(\gP^{*}, y_{0})$ and all $c>0$, with probability at least $1-e^{-x_{0}}$,
\begin{align*}
\left|\gZ(\gX, \bsg_{1}, \bsg_{2})\right|\leq&\ \cZ^{\sbP_{m_{1}}\times\sbP_{m_{2}}}(\gX)\\
\leq&\ (1+c)\left(9.77\sqrt{\frac{2\alpha+1/2}{\eta}}+\frac{90(2\alpha+1/2)}{\eta}\right)y_{0}^{2}\\&+\cro{\frac{3\sqrt{2}c}{16}+\frac{4}{a}\left(1+\frac{8}{c}\right)}y_{0}^{2}.
\end{align*}
To minimize the bracketed term, we take $$c=\sqrt{\frac{32}{\left(9.77\sqrt{\frac{2\alpha+1/2}{\eta}}+\frac{90(2\alpha+1/2)}{\eta}+\frac{3\sqrt{2}}{16}\right)a}}$$ and therefore for all $(\gP_{\bsg_{1}},\gP_{\bsg_{2}})\in\sbB^{\sbP_{m_{1}}\times\sbP_{m_{2}}}(\gP^{*}, y_{0})$ with probability at least $1-e^{-x_{0}}$,
\[
\left|\gZ(\gX, \bsg_{1}, \bsg_{2})\right|\leq\cZ^{\sbP_{m_{1}}\times\sbP_{m_{2}}}(\gX)\leq ak_{0}x_{0}.
\]
Combining all the bounds together, we derive for all $(\gP_{\bsg_{1}},\gP_{\bsg_{2}})\in\sbP_{m_{1}}\times\sbP_{m_{2}}$ simultaneously with probability at least $1-\varepsilon(m_{1},m_{2})$,
\[
\left|\gZ(\gX, \bsg_{1}, \bsg_{2})\right|\leq k_{1}\cro{\gh^{2}(\gP^{*},\gP_{\bsg_{1}})+\gh^{2}(\gP^{*},\gP_{\bsg_{2}})}+ak_{0}x_{0}(m_{1},m_{2}),
\]
where $$\varepsilon(m_{1},m_{2})=2\exp\cro{-x_{0}(m_{1},m_{2})}+\sum_{j\geq1}\exp\cro{-x_{j}(m_{1},m_{2})}.$$ By the definition of $x_{j}(m_{1},m_{2})$, we notice that for all $j\in\N$, $x_{j}(m_{1},m_{2})\geq\Delta(m_{1})+\Delta(m_{2})+\vartheta\delta^{j}+\xi$. Hence, provided (\ref{delta-upsilon-bound}), we have
\begin{align*}
\varepsilon(m_{1},m_{2})&\leq\exp\cro{-\xi-\Delta(m_{1})-\Delta(m_{2})}\left(2\exp(-\vartheta)+\sum_{j\geq1}\exp(-\vartheta\delta^{j})\right)\\
&\leq\exp\cro{-\xi-\Delta(m_{1})-\Delta(m_{2})}.
\end{align*}
Finally we can extend this result to all $(\gP_{\bsg_{1}},\gP_{\bsg_{2}})\in\widetilde\sbP\times\widetilde\sbP$ by summing these bounds over $(m_{1},m_{2})\in\cM\times\cM$ and using \eref{weight}.
\end{proof}
\begin{proof}[Proof of Theorem~\ref{thm-1}]
We apply Proposition~\ref{empirical-process-bound} with $\delta=1.175$, $\vartheta=1.47$ and as for the values of $\eta$ and $a$, we shall choose them later such that $k_{1}=3\beta/8$, with some $0<\beta<1$. On a set $\Omega_{\xi}$ the probability of which is at least $1-\Sigma^{2}e^{-\xi}$, for all $\gP_{\bsg_{1}},\gP_{\bsg_{2}}\in\widetilde\sbP$ and all $\sbP_{m_{1}}\times\sbP_{m_{2}}$ containing $(\gP_{\bsg_{1}},\gP_{\bsg_{2}})$
\begin{align*}
\gT(\gX,\bsg_{1},\bsg_{2})\leq&\ \E\cro{\gT(\gX,\bsg_{1},\bsg_{2})}+\frac{3\beta}{8}\cro{\gh^{2}(\gP^{*},\gP_{\bsg_{1}})+\gh^{2}(\gP^{*},\gP_{\bsg_{2}})}\\
&+k_{0}a\cro{\eta\left(D_{n}(m_{1})+D_{n}(m_{2})\right)\vee\left(\Delta(m_{1})+\Delta(m_{2})+\vartheta+\xi\right)}\\
\leq&\ \E\cro{\gT(\gX,\bsg_{1},\bsg_{2})}+\frac{3\beta}{8}\cro{\gh^{2}(\gP^{*},\gP_{\bsg_{1}})+\gh^{2}(\gP^{*},\gP_{\bsg_{2}})}\\
&+k_{0}a\cro{\eta D_{n}(m_{1})+\eta D_{n}(m_{2})+\Delta(m_{1})+\Delta(m_{2})+\vartheta+\xi}.
\end{align*}
Since the last inequality is true for all the $\sbP_{m_{1}}\times\sbP_{m_{2}}$ containing $(\gP_{\bsg_{1}},\gP_{\bsg_{2}})$, provided $C_{0}(2\alpha+1/2)\geq k_{0}a\eta$, we derive from \eref{penalty-procedure} that with a probability at least $1-\Sigma^{2}e^{-\xi}$,
\begin{align}
\gT(\gX,\bsg_{1},\bsg_{2})\leq&\ \E\cro{\gT(\gX,\bsg_{1},\bsg_{2})}+\frac{3\beta}{8}\cro{\gh^{2}(\gP^{*},\gP_{\bsg_{1}})+\gh^{2}(\gP^{*},\gP_{\bsg_{2}})}\notag\\
&+\gpen(\bsg_{1})+\gpen(\bsg_{2})+k_{0}a(\vartheta+\xi).\label{T-bound-1}
\end{align}
According to Proposition~3 of \cite{BarBir2018}, the function $\psi$ satisfies Assumption~2 (more precisely (10)) in the same paper with $a_{0}=4$ and $a_{1}=3/8$. As a consequence, for all $\gP_{\bsg_{1}},\gP_{\bsg_{2}}\in\widetilde\sbP$ and $\gP^{*}\in\sbP$,
\begin{equation}\label{expec-bound}
\E\cro{\gT(\gX,\bsg_{1},\bsg_{2})}\leq4\gh^{2}(\gP^{*},\gP_{\bsg_{1}})-\frac{3}{8}\gh^{2}(\gP^{*},\gP_{\bsg_{2}}).
\end{equation}
Combining \eref{T-bound-1} and \eref{expec-bound}, we derive that for all $\gP_{\bsg_{1}},\gP_{\bsg_{2}}\in\widetilde\sbP$ and $\gP^{*}\in\sbP$, with a probability at least $1-\Sigma^{2}e^{-\xi}$,
\begin{align}
\gT(\gX,\bsg_{1},\bsg_{2})\leq&\ (4+\frac{3\beta}{8})\gh^{2}(\gP^{*},\gP_{\bsg_{1}})-\frac{3(1-\beta)}{8}\gh^{2}(\gP^{*},\gP_{\bsg_{2}})\nonumber\\
&+\gpen(\bsg_{1})+\gpen(\bsg_{2})+k_{0}a(\vartheta+\xi).\label{T-bound-2}
\end{align}
This entails that, for any (random) elements $\gP_{\widehat\bsg_{\lambda}}, \gP_{\widehat\bsg_{\widehat\lambda}}\in\widetilde\sbP$, on a set $\Omega_{\xi}$ with probability at least $1-\Sigma^{2}e^{-\xi}$
\begin{align}
\gT(\gX,\widehat\bsg_{\lambda},\widehat\bsg_{\widehat\lambda})\leq&\ (4+\frac{3\beta}{8})\gh^{2}(\gP^{*},\gP_{\widehat\bsg_{\lambda}})-\frac{3(1-\beta)}{8}\gh^{2}(\gP^{*},\gP_{\widehat\bsg_{\widehat\lambda}})\label{T-bound-3}\\
&+\gpen(\widehat\bsg_{\lambda})+\gpen(\widehat\bsg_{\widehat\lambda})+k_{0}a(\vartheta+\xi)\notag
\end{align}
and
\begin{align}
\gup(\bsX,\widehat\bsg_{\lambda})=&\ \sup_{\lambda'\in\Lambda}\cro{\gT(\bsX,\widehat\bsg_{\lambda},\widehat\bsg_{\lambda'})-\gpen(\widehat\bsg_{\lambda'})}+\gpen(\widehat \bsg_{\lambda})\notag\\
\leq&\ (4+\frac{3\beta}{8})\gh^{2}(\gP^{*},\gP_{\widehat\bsg_{\lambda}})-\frac{3(1-\beta)}{8}\inf_{\lambda'\in\Lambda}\gh^{2}(\gP^{*},\gP_{\widehat\bsg_{\lambda'}})\label{T-bound-4}\\
&+2\gpen(\widehat\bsg_{\lambda})+k_{0}a(\vartheta+\xi).\notag
\end{align}
By the construction of $\psi$, $\gT(\gX,\widehat\bsg_{\widehat\lambda},\widehat\bsg_{\lambda})=-\gT(\gX,\widehat\bsg_{\lambda},\widehat\bsg_{\widehat\lambda})$. Combining \eref{T-bound-3}, \eref{T-bound-4} and \eref{def-sE-2} leads to for any $\lambda\in\Lambda$, on a set $\Omega_{\xi}$ with probability at least $1-\Sigma^{2}e^{-\xi}$
\begin{align}
\frac{3(1-\beta)}{8}\gh^{2}(\gP^{*},\gP_{\widehat\bsg_{\widehat\lambda}})\leq&\ (4+\frac{3\beta}{8})\gh^{2}(\gP^{*},\gP_{\widehat\bsg_{\lambda}})-\gT(\gX,\widehat\bsg_{\lambda},\widehat\bsg_{\widehat\lambda})\notag\\
&+\gpen(\widehat\bsg_{\lambda})+\gpen(\widehat\bsg_{\widehat\lambda})+k_{0}a(\vartheta+\xi)\notag\\
\leq&\ (4+\frac{3\beta}{8})\gh^{2}(\gP^{*},\gP_{\widehat\bsg_{\lambda}})+\cro{\gT(\gX,\widehat\bsg_{\widehat\lambda},\widehat\bsg_{\lambda})-\gpen(\widehat\bsg_{\lambda})}\notag\\
&+\gpen(\widehat\bsg_{\widehat\lambda})+2\gpen(\widehat\bsg_{\lambda})+k_{0}a(\vartheta+\xi)\notag\\
\leq&\ (4+\frac{3\beta}{8})\gh^{2}(\gP^{*},\gP_{\widehat\bsg_{\lambda}})+\gup(\bsX,\widehat\bsg_{\widehat\lambda})\notag\\&+2\gpen(\widehat\bsg_{\lambda})+k_{0}a(\vartheta+\xi)\notag\\
\leq&\ (4+\frac{3\beta}{8})\gh^{2}(\gP^{*},\gP_{\widehat\bsg_{\lambda}})+\gup(\bsX,\widehat\bsg_{\lambda})+1\notag\\&+2\gpen(\widehat\bsg_{\lambda})+k_{0}a(\vartheta+\xi).\label{T-bound-5}
\end{align}
Plugging (\ref{T-bound-4}) into (\ref{T-bound-5}) yields, for any $\lambda\in\Lambda$, on a set $\Omega_{\xi}$ with probability at least $1-\Sigma^{2}e^{-\xi}$,
\[
\frac{3(1-\beta)}{8}\gh^{2}(\gP^{*},\gP_{\widehat\bsg_{\widehat\lambda}})\leq(8+\frac{3\beta}{4})\gh^{2}(\gP^{*},\gP_{\widehat\bsg_{\lambda}})+4\gpen(\widehat\bsg_{\lambda})+2k_{0}a(\vartheta+\xi)+1.
\]
Therefore, for any $\lambda\in\Lambda$ on a set $\Omega_{\xi}$ with probability at least $1-\Sigma^{2}e^{-\xi}$,
\begin{equation}\label{p-s}
\gh^{2}(\gP^{*},\gP_{\widehat\bsg_{\widehat\lambda}})\leq\frac{64+6\beta}{3(1-\beta)}\gh^{2}(\gP^{*},\gP_{\widehat\bsg_{\lambda}})+\frac{32}{3(1-\beta)}\gpen(\widehat\bsg_{\lambda})+\frac{16k_{0}a(\vartheta+\xi)+8}{3(1-\beta)}.
\end{equation}
By the equality \eref{l3}, we rewrite \eref{p-s} as the following
\begin{equation}\label{T-bound-6}
\gh^{2}(\gR^{*},\gR_{\widehat\bsg_{\widehat\lambda}})\leq\frac{64+6\beta}{3(1-\beta)}\gh^{2}(\gR^{*},\gR_{\widehat\bsg_{\lambda}})+\frac{32}{3(1-\beta)}\gpen(\widehat\bsg_{\lambda})+\frac{16k_{0}a(\vartheta+\xi)+8}{3(1-\beta)}.
\end{equation}
Taking $\beta=0.75$, $\eta\approx9947.13(2\alpha+1/2)$, we can compute the value of $a\approx2365.57$ such that $k_{1}=3\beta/8$ and $k_{0}\approx0.251$. Therefore, provided $C_{0}\geq5.9\times10^{6}$, plugging the values of $\beta$, $k_{0}$, $a$ and $\vartheta$ into \eref{T-bound-6}, we finally conclude.
\end{proof}
\section{Signals for testing Poisson and exponential models} 

\texttt{fms-type} (Poisson): $n=497$, changepoints are located at the positions 
\begin{equation*}
l_{0}=\left(\frac{139}{497}, \frac{226}{497}, \frac{243}{497}, \frac{300}{497}, \frac{309}{497}, \frac{333}{497}\right). 
\end{equation*}
The Poisson mean on each segment is 4, 6, 10, 3, 7, 1, 5 respectively, i.e. $\bsg^{*}$ takes the value $\log 4$, $\log 6$, $\log 10$, $\log3$, $\log7$, $\log1$, $\log5$ on each segment. For this signal, we also test the scenario when outliers present in the observations by randomly modifying five points in the observations into 30.

\texttt{mix-type} (Poisson): $n=560$ and $\bsg^{*}$ is a piecewise constant function on $[0,1)$ with 13 changepoints at a sequence of locations 
\begin{equation*}
l_{0}=\left(\frac{11}{560}, \frac{21}{560}, \frac{41}{560}, \frac{61}{560}, \frac{91}{560}, \frac{121}{560}, \frac{161}{560}, \frac{201}{560}, \frac{251}{560}, \frac{301}{560}, \frac{361}{560}, \frac{421}{560}, \frac{491}{560}\right)
\end{equation*}
and on each segment the Poisson mean $e^{\bsg^{*}}$ is given by the value 30, 2, 26, 4, 24, 6, 22, 8, 20, 10, 18, 12, 16, 14 respectively.

\texttt{teeth-type} (exponential): $n=140$ and $\bsg^{*}$ is a piecewise constant function on $[0,1)$ with 13 changepoints at a sequence of locations 
\begin{equation*}
l_{0}=\left(\frac{11}{140}, \frac{21}{140}, \frac{31}{140}, \frac{41}{140}, \frac{51}{140}, \frac{61}{140}, \frac{71}{140}, \frac{81}{140}, \frac{91}{140}, \frac{101}{140}, \frac{111}{140}, \frac{121}{140}, \frac{131}{140}\right)
\end{equation*}
and on each segment the value of $\bsg^{*}$ is given by 0.5, 5, 0.5, 5, 0.5, 5, 0.5, 5, 0.5, 5, 0.5, 5, 0.5, 5 respectively. For this signal, we also test the scenario when outliers present in the observations by randomly modifying two points in the observations into 20.

\texttt{stairs-type} (exponential): $n=500$ and $\bsg^{*}$ is a piecewise constant function on $[0,1)$ with 4 changepoints at a sequence of locations 
\begin{equation*}
l_{0}=\left(\frac{101}{500}, \frac{201}{500}, \frac{301}{500}, \frac{401}{500}\right)
\end{equation*}
and on each segment the value of $\bsg^{*}$ is given by $2^{4}$, $2^{2}$, 1, $2^{-2}$, $2^{-4}$ respectively.

\end{document}